\documentclass[12pt]{iopart}
\usepackage{iopams}
\usepackage{epsf}
\usepackage{latexsym}
\usepackage{epsfig}
\usepackage{amssymb}
\usepackage{amsfonts}
\usepackage{graphicx}

\begin{document}

\title{Effects of hidden nodes on network structure inference}

\author{Haiping Huang$^{1,2}$}
\address{$^{1}$ RIKEN Brain Science Institute, Wako-shi, Saitama
351-0198, Japan}
\address{$^{2}$ Department of Computational Intelligence and Systems Science,
Tokyo Institute of Technology, Yokohama 226-8502, Japan}
\date{\today}

\begin{abstract}
 Effects of hidden nodes on inference quality of observed network
 structure are explored based on a disordered Ising model with hidden
 nodes. We first study analytically small systems consisting of a
 few nodes, and find that the magnitude of the effective coupling grows as the coupling strength from the hidden
 common input nodes increases, while the field strength of the input node
 has opposite effects. Insights gained from analytic results of
 small systems are confirmed in numerical simulations of large
 systems. We also find that the inference quality deteriorates as
 the number of hidden nodes increases. Furthermore, increasing field variance of
hidden nodes improves the inference quality of the effective
couplings, but worsens the quality for the effective fields. In
addition, attenuated coupling strengths involved in at least one
hidden node lead to high quality of coupling inference.

\end{abstract}

 \maketitle

\section{Introduction}
Due to recent progresses in multi-electrode recording techniques in
experimental neuroscience, the neural activity measurement of a
population of increasing number of neurons
becomes possible~\cite{Marre-2012,Buzaki-2014}, which provides a large
opportunity and also a big challenge for the large-scale data
analysis or dimensionality reduction~\cite{Yu-2014}, to understand
how sensory processing, working memory or decision making arises
from neural circuits. However, the brain region the current
recording techniques can measure is very limited, therefore, the
measured population of neurons is not completely isolated. Still
there are many unobserved neurons outside this population (some may
be upstream neurons), but interacting with neurons inside the
observed population~\cite{Rieke-2008, Hertz-11, Tkacik-2014,
Paninski-13}. The inferred interactions from the neural data are
usually termed functional or effective connectivity, since the data
fitting captures only part of the statistical features of the spike
train data for algorithmic simplicity.

Recently, there appear intensive research interests on the inference
problem of a random kinetic Ising model with hidden
nodes~\cite{Friedman-1998,Roudi-2013,Hertz-2014,Opper-2014}.
However, a systematic study of the effects of hidden nodes on the
structure inference is still lacking based on widely used
equilibrium models. Here, we explore these effects based on the
maximal entropy model (also called the disordered Ising model)
extensively used to model and fit the neural data in recent
studies~\cite{Nature-06,Huang-2012pre,Cocco-13,Tkacik-2014}. We
separate an unobserved part of network (a hidden subnetwork) from
the full model, and probe the effects of hidden nodes first in small
systems with a few nodes where the analytic study is available, then
in large systems where we performed extensive numerical simulations
under various settings. In fact, the correlation observed between
two neurons may arise from a hidden common input outside the
measured population. In addition, the interaction between two
neurons may be mediated by a chain of unobserved neurons. How the
unobserved part of the network affects the final inference quality
will be the focus of this paper.

For a model study, the data used for inference are collected with
high quality, such that the final result does not strongly depend on
the inference method. In numerical simulations, we used the naive
mean-field (nMF) method~\cite{Kappen-1998,Tanaka-1998,Huang-2013} to
reconstruct the interaction strengths between neurons in the
observed network. nMF relies simply on the inverse of the
correlation matrix, which is also suitable for analytic studies of
small systems with a few nodes. For small systems, we found that the
effective field of observed neurons receiving a common input increases with
the external field (firing bias) applied to the hidden input, while neurons interacting directly with the hidden input show
different behavior from those that do not have a direct
interaction. The effective couplings between neurons receiving a common input
decreases with increasing firing bias of the hidden input. However, the
neurons without a direct interaction (with the hidden input) seem to
be unaffected in terms of their inferred coupling values.
Furthermore, if two neurons are mediated by a chain of hidden neurons, the
inferred interaction strength decreases with increasing chain
length. Insights gained from the small size network are confirmed in
numerical simulations of large size networks, from which, we showed
that the inference performance deteriorates with increasing size of the
unobserved network, and enhanced coupling strengthes between neurons
in the unobserved part will increase the inference error for both
the couplings and fields. Interestingly, if we increase the field
strength of the unobserved neurons, the inference of couplings for
the observed neurons will be improved, while the quality of
the field inference still deteriorates.

The rest of this paper is organized as follows. The disordered Ising
model compatible with the data is defined in Sec.~\ref{sec_Ising}.
In Sec.~\ref{sec_ana}, we present an analytic study on small
systems. In Sec.~\ref{sec:num}, we present numerical simulation
results on large systems. Conclusion is given in Sec.~\ref{sec_con}.

\section{Disordered Ising model with hidden nodes}
\label{sec_Ising} The experimental data can be described by $P$
independent sampled configurations
$\{\boldsymbol{\sigma}^{\mu}\}(\mu=1,\ldots,P)$ where
$\boldsymbol{\sigma}$ is an $N-$dimensional vector (the entry of the
vector takes a binary value $\pm1$) and $N$ is the network size. To
build a minimal model to fit these data, one can take the
constraints up to the second-order correlations in the data,
resulting in the following maximal entropy model~\cite{Jaynes-1957}:
\begin{equation}\label{MaxEnt}
    P(\boldsymbol{\sigma})=\frac{1}{Z}\exp\left[\sum_{i<j}J_{ij}\sigma_i\sigma_j+\sum_{i}h_i\sigma_i\right].
\end{equation}
The coupling terms $\{J_{ij}\}$ correspond to the correlation
constraints
($\left<\sigma_i\sigma_j\right>_{P(\boldsymbol{\sigma})}=\left<\sigma_i\sigma_j\right>_{{\rm
data}}$), while the field terms $\{h_i\}$ correspond to the
magnetization constraints
($\left<\sigma_i\right>_{P(\boldsymbol{\sigma})}=\left<\sigma_i\right>_{{\rm
data}}$). $Z$ is a normalization constant. The symmetric coupling
may take either positive value or negative value, and the field also
has the same situation, depending on the
data~\cite{Huang-2012pre,Tkacik-2014}. Hence, in general, we call
this data-driven model a disordered Ising model or a spin glass
model~\cite{Mezard-1987}.

From Eq.~(\ref{MaxEnt}), one can define an energy term
$E=-\sum_{i<j}J_{ij}\sigma_i\sigma_j-\sum_{i}h_i\sigma_i$, then the
distribution in Eq.~(\ref{MaxEnt}) is known as a Boltzmann
distribution in statistical mechanics~\cite{SM-1992}. In this paper,
we divide the full network into observed (visible) part and
unobserved (hidden) part, and study the effects of hidden part on
the inference quality. We also assume the observed neurons have
originally no firing biases. Therefore, the energy term becomes,
\begin{equation}\label{hidden}
   E=-\sum_{i<j}J_{ij}^{ab}\sigma_i^{a}\sigma_j^{b}-\sum_{i}h_i^{H}\sigma_i^{H},
\end{equation}
where $a,b=V,H$, and $J_{ij}^{VV}$ $(J_{ij}^{HH})$ specifies the
coupling strength between two neurons from the visible (hidden)
part, and $J_{ij}^{VH}$ $(J_{ij}^{HV})$ indicates the coupling
strength between two neurons one of which comes from the hidden
part. All these coupling strengthes follow a Gaussian distribution
with zero mean and variance $\sigma_J/c$. The field also follows a
Gaussian distribution with zero mean and variance $\sigma_h$. In the
simulation, one can also control the number of neurons in the hidden
part, defined by $N_h$. In the full network, each neuron is
connected to $c$ other neurons on average without self-interaction.

In general, it is quite difficult to infer the couplings or fields
related to hidden nodes given the observed
data~\cite{Ack-1985,Kappen-1998}, although some special connection
structure could be assumed for hidden part to make the inference
possible at large network sizes~\cite{Saul-1996,Hinton-2006}. Here,
we focus on the effects of hidden part on the inference quality of
network structure of observed part. For simplicity, we used naive
mean-field method to infer the couplings and fields in observed part
and tested the inference performance against various control
parameters. First, we define a connected correlation
$C_{ij}=\left<\sigma_i\sigma_j\right>_{{\rm data}}-m_im_j$, where
the magnetization $m_i=\left<\sigma_i\right>_{\rm data}$. Then the
coupling between neuron $i$ and neuron $j$ can be reconstructed
by~\cite{Kappen-1998}
\begin{equation}\label{nMFJ}
    J_{ij}^{VV}=-(\mathbf{C}^{-1})_{ij}.
\end{equation}
After the coupling is obtained, the field for neuron $i$ is inferred
by~\cite{Kappen-1998}
\begin{equation}\label{nMFh}
    h_{i}^{V}=\tanh^{-1}(m_i)-\sum_{j\neq i}J_{ij}^{VV}m_j-m_i\left[\frac{1}{1-m_{i}^{2}}-(\mathbf{C}^{-1})_{ii}\right].
\end{equation}
Eq.~(\ref{nMFJ}) is derived based on the mean-field equation
$m_i=\tanh(\sum_{j\neq i}J_{ij}m_j+h_i)$ and the linear-response
theory $C_{ij}=\partial m_i/\partial
h_j$~\cite{Kappen-1998,Tanaka-1998}. A variety of advanced mean
field approximations can be reduced to this simple approximation
under certain conditions, e.g., high temperature or small correlation~\cite{Huang-2013}. The last term in
Eq.~(\ref{nMFh}) is related to an effective self-coupling playing a key
role in accurate field inference. However, a more natural and
accurate way for field inference is applying an adaptive Onsager
correction term~\cite{Huang-2013}. For simplicity, we used the naive
mean-field method in this paper.

\begin{figure}
\centering
    \includegraphics[bb=90 643 481 736,width=0.8\textwidth]{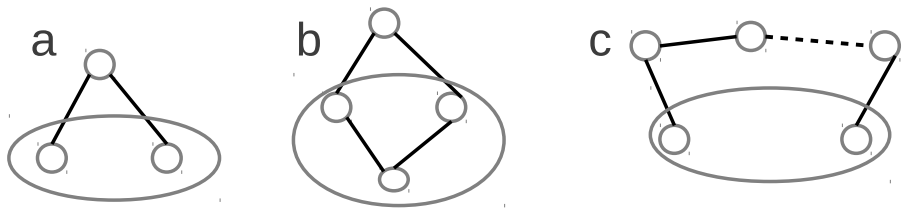}
  \caption{
  Small systems with hidden nodes. Neurons are
  indicated by gray circles, and the line indicates the direct
  interaction. The observed part is marked by an ellipse. Two neurons receiving a hidden
  common input (a). Three neurons with only two of them receiving a
  hidden common input (b). The interaction between two neurons
  mediated by a chain of unobserved neurons (c). The dashed line
  indicates the other $(L-3)$ unobserved neurons.
  }\label{Small}
\end{figure}

\section{Small size system: analytic studies}
\label{sec_ana}

In this section, we study analytically effects of hidden nodes on
small size system, which could further provide insights towards
understanding these effects on large systems.

\subsection{Three-neuron system with one common hidden input}
\label{subsec:ana01}

We first consider a three-neuron system with one common hidden input
shown in Fig.~\ref{Small} (a). We assume the observed neurons
interact with the hidden neuron with the same coupling strength $J$,
and the hidden neuron has a firing bias $h$. The normalization
constant $Z_{a}=2\sum_{\sigma,\sigma'}\cosh(J\sigma+J\sigma'+h)$
where ${\sigma,\sigma'}$ are states of the observed neurons.
Therefore, the exact observed connected correlation ($C$) are given
by:
\begin{eqnarray}\label{case-a}
C&=&1-\frac{4\cosh h}{\cosh(2J+h)+\cosh(2J-h)+2\cosh h}-mm',\\
m&=&m'=\frac{\cosh(2J+h)-\cosh(2J-h)}{\cosh(2J+h)+\cosh(2J-h)+2\cosh
h},
\end{eqnarray}
where $m(m')$ is the observed magnetization. According to
Eq.~(\ref{nMFJ}), the inferred coupling is given by:
\begin{equation}\label{J-a}
    J_{{\rm eff}}=\frac{C}{(1-m^{2})(1-m'^{2})-C^{2}}.
\end{equation}
The field can be predicted by:
\begin{equation}\label{h-a}
   h_{\rm eff}=h'_{{\rm eff}}=\tanh^{-1}(m)-J_{{\rm eff}}m-
   m\left[\frac{1}{1-m^{2}}-\frac{1-m^{2}}{(1-m^{2})^{2}-C^{2}}\right].
\end{equation}
If we increase the field $h$ to a very large value, then we can get the correlation difference $\Delta C=C(h=0)-C(h\rightarrow\infty)=m^2$ and the
coupling difference $\Delta J_{\rm{eff}}=\frac{m^2}{1-m^4}$, where $m=\frac{\sinh 2J}{1+\cosh 2J}$. Therefore, both differences are positive,
implying that increasing the field will lower down both the correlations and the inferred couplings.

\begin{figure}[h!]
 (a) \includegraphics[bb=53 25 719 518,scale=0.33]{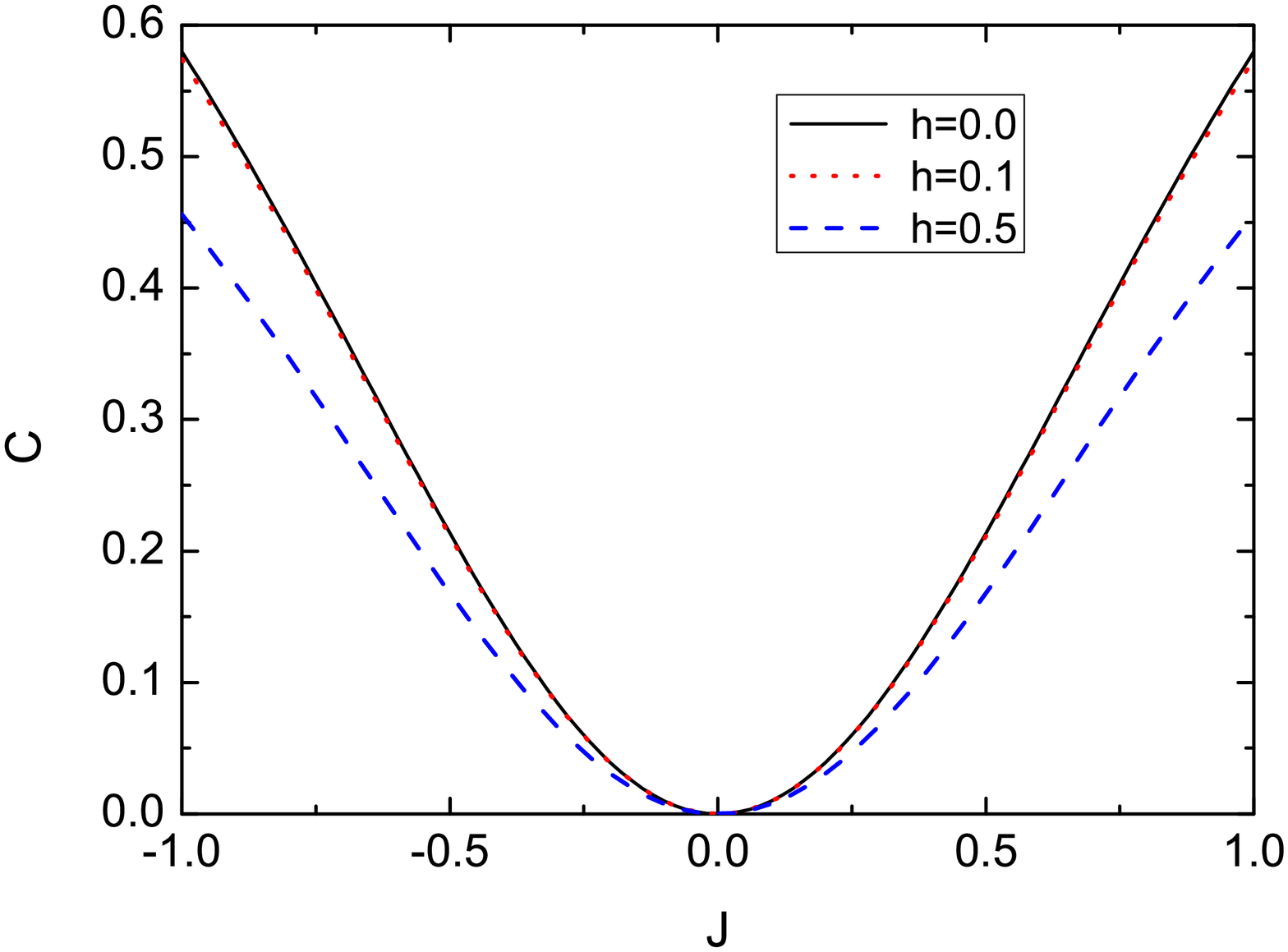}
     \hskip .2cm
  (b)   \includegraphics[bb=56 25 717 519,scale=0.33]{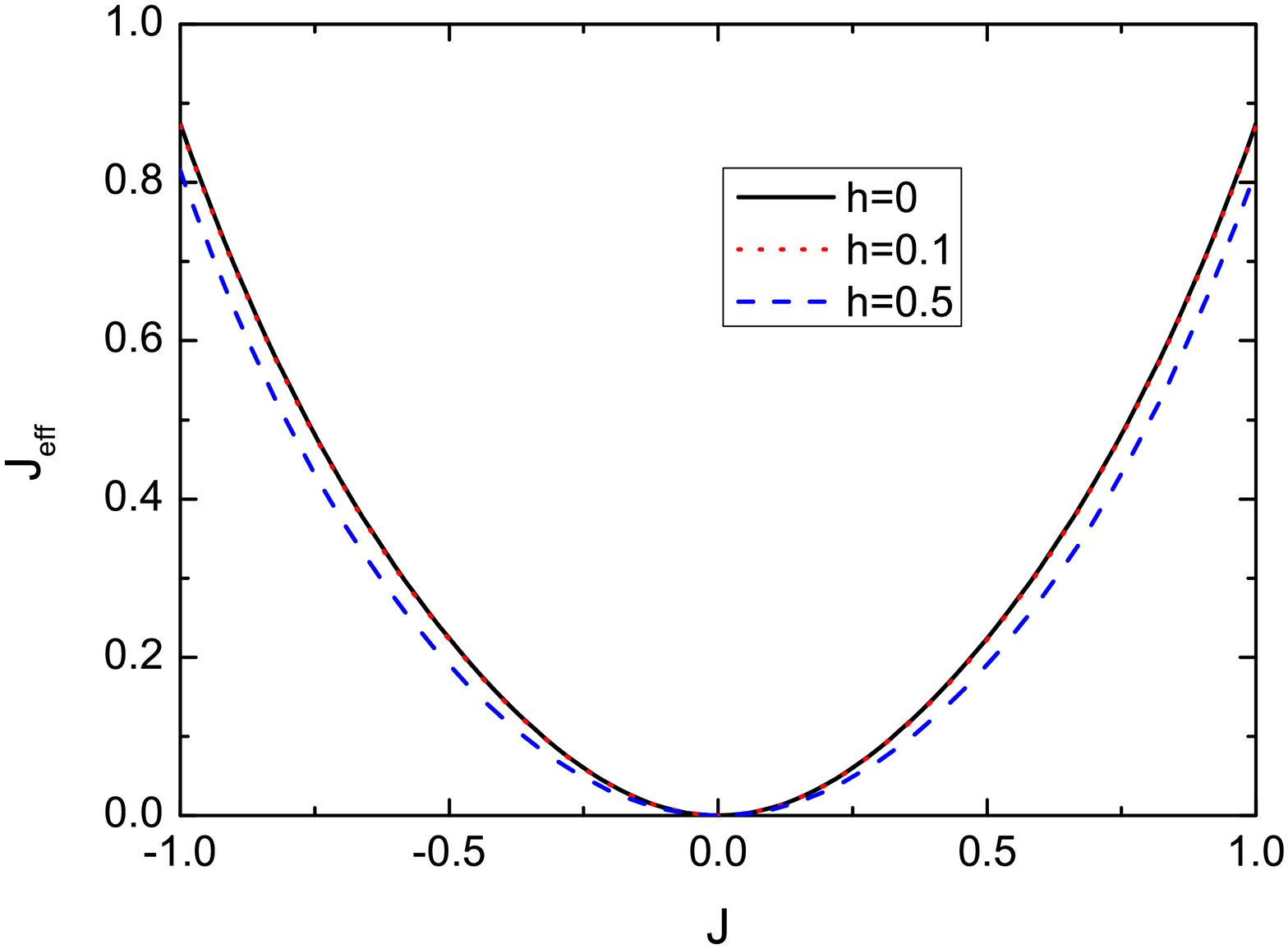}
     \vskip .2cm
     \centering
   (c)  \includegraphics[bb=53 25 713 511,scale=0.4]{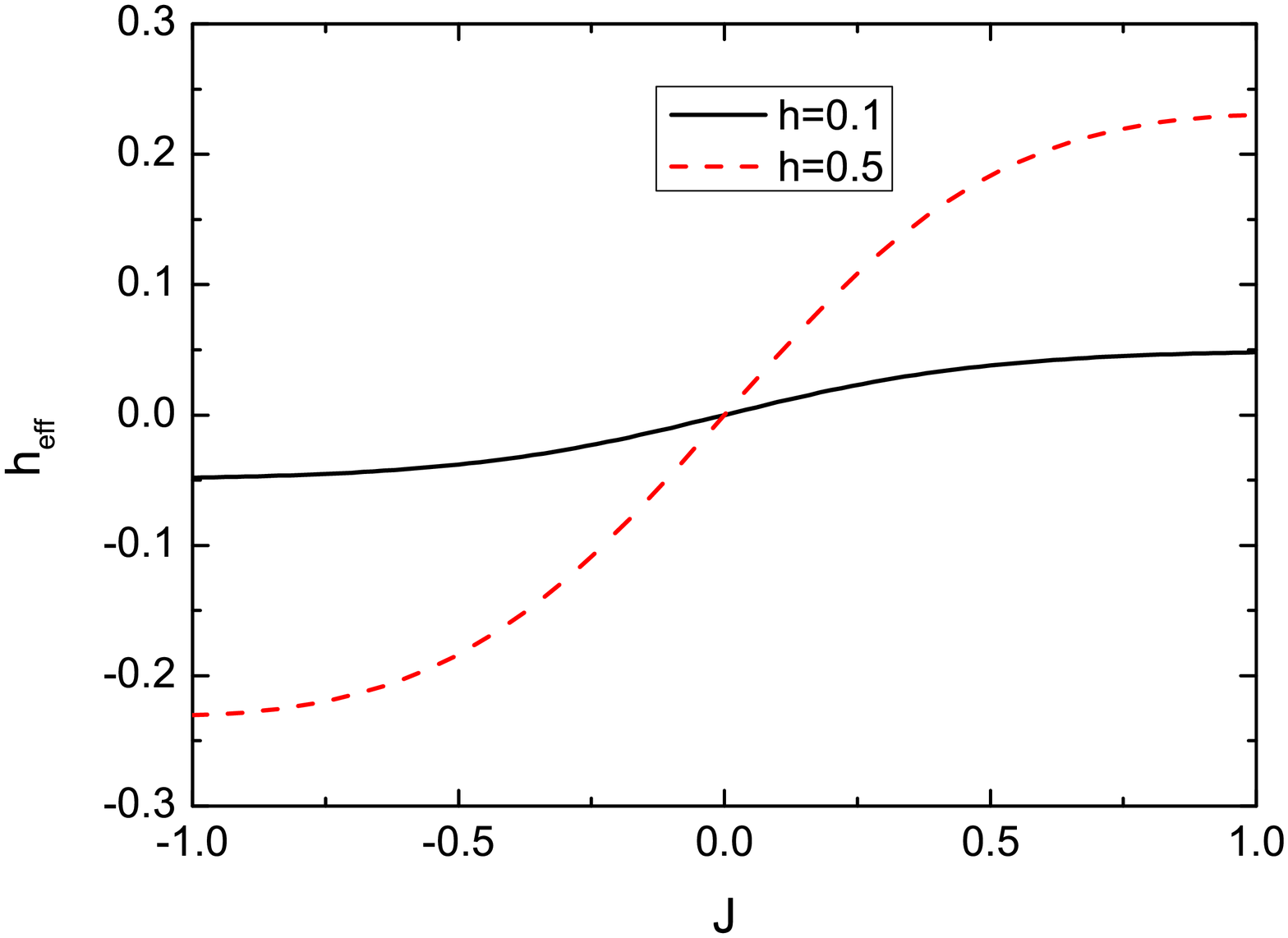}\vskip .2cm
  \caption{(Color online)
     Inference results corresponding to Fig.~\ref{Small} (a). (a)
     connected correlation versus coupling strength between observed
     neurons and the hidden one. (b) effective coupling between observed
     neurons. Note that the true value is zero. (c) effective field
     for observed neurons. Note also that the true value is zero.
   }\label{ana01}
 \end{figure}

Results are shown in Fig.~\ref{ana01}. As observed in
Fig.~\ref{ana01} (a), increasing the external field of hidden input
will lower down the connected correlation between two observed
neurons. This effect also makes the predicted coupling between two
observed neurons smaller than that in the presence of a smaller external field, as
shown in Fig.~\ref{ana01} (b). However, the effective coupling
increases as the absolute value of the coupling strength
($J^{VH}=J$) grows. Only when $J$ becomes very weak, $J_{\rm eff}$
gets close to zero which is the true value, due to the nature of the
naive mean field approximation. The behavior of effective fields
with $J$ and $h$ is shown in Fig.~\ref{ana01} (c). The external
field of hidden input will increase the magnitude of the effective
field applied on the observed neurons. For large $J$, the effective field seems to
saturate to some large values but smaller than $h$. When $J$ changes
its sign, so does $h_{{\rm eff}}$, which can be zero only when no
hidden input interacts with the observed part.

\subsection{Four-neuron system with one common hidden input}
\label{subsec:ana02} We consider one additional neuron in the
observed part, but it doses not interact directly with the hidden
neuron, as shown in Fig.~\ref{Small} (b). We assume all $J^{VH}=J$,
$J^{VV}=J$ except for the neurons interacting with the hidden
neuron ($J_{12}^{VV}=0$). In the observed part, three neurons are given the indexes
$1,2,3$. The hidden neuron has a firing bias
$h$. In this case, $Z_b=4\cosh(2J)(\cosh(2J+h)+\cosh(2J-h))+8\cosh
h$. The connected correlation between neuron $1$ and neuron $2$ is
given by:
\begin{eqnarray}\label{case-b}
C&=&\frac{\cosh(2J)(\cosh(2J+h)+\cosh(2J-h))-2\cosh h}
{\cosh(2J)(\cosh(2J+h)+\cosh(2J-h))+2\cosh h}-m_1m_2,\\
m_1&=&m_2=\frac{\cosh(2J)(\cosh(2J+h)-\cosh(2J-h))}{\cosh(2J)(\cosh(2J+h)+\cosh(2J-h))+2\cosh
h}.
\end{eqnarray}
$C_{13}$ and $C_{23}$ are given by:
\begin{eqnarray}\label{case-b02}
\fl C_{13}=C_{23}=\frac{\sinh(2J)(\cosh(2J+h)+\cosh(2J-h))}
{\cosh(2J)(\cosh(2J+h)+\cosh(2J-h))+2\cosh h}-m_1m_3,\\
\fl
m_3=\frac{\sinh(2J)(\cosh(2J+h)-\cosh(2J-h))}{\cosh(2J)(\cosh(2J+h)+\cosh(2J-h))+2\cosh
h}.
\end{eqnarray}
Let $m_1=m_2=m$, $m_3=m'$, $C_{13}=C_{23}=C'$, we have the following
inferred results:
\begin{eqnarray}\label{J-b}
 J_{{\rm eff}}&=&\frac{C(1-m'^{2})-C'^{2}}
{\Delta},\\
 J_{{\rm in}}&=&\frac{C'(1-m^{2}-C)}{\Delta},
\end{eqnarray}
where $J_{{\rm eff}}$ and $J_{{\rm in}}$ are inferred values of $J_{12}$ and
$J_{23}(J_{13})$ respectively.
$\Delta=(1-m^{2})((1-m^{2})(1-m'^{2})-C'^{2})-C(C(1-m'^{2})-C'^{2})+C'^{2}(C-1+m^{2})$. It is easy to show that
$C(h=0)-C(h\rightarrow\infty)=(\cosh^2(2J)-1)\cosh^2(2J)/(1+\cosh^2(2J))^2\ge0$ and $|C'(h=0)|-|C'(h\rightarrow\infty)|=\cosh(2J)|\sinh(2J)|(\cosh^2(2J)-1)/(1+\cosh^2(2J))^2\ge0$.
One can also show that $J_{{\rm eff}}(h=0)-J_{{\rm eff}}(h\rightarrow\infty)=(\cosh^4(2J)-1)/(8\cosh^2(2J))\ge0$ and $J_{{\rm in}}(h=0)=J_{{\rm in}}(h\rightarrow\infty)=
(\cosh^2(2J)+1)\tanh(2J)/4$. These results can explain the interesting properties shown below.
The fields are inferred as:
\begin{eqnarray}\label{h-b}
\fl  h_{{\rm eff}}&=&\tanh^{-1}(m)-J_{{\rm eff}}m-J_{{\rm
in}}m'-m\left[\frac{1}{1-m^{2}}-\frac{(1-m^{2})(1-m'^{2})-C'^{2}}
{\Delta}\right],\\
\fl h_{{\rm in}}&=&\tanh^{-1}(m')-2J_{{\rm
in}}m-m'\left[\frac{1}{1-m'^{2}}-\frac{(1-m^{2})^{2}-C^{2}}
{\Delta}\right],
\end{eqnarray}
where $h_{{\rm eff}}$ and $h_{{\rm in}}$ are the inferred values of $h_1(h_2)$
and $h_3$ respectively.

\begin{figure}[h!]
 (a) \includegraphics[bb=64 23 714 517,scale=0.33]{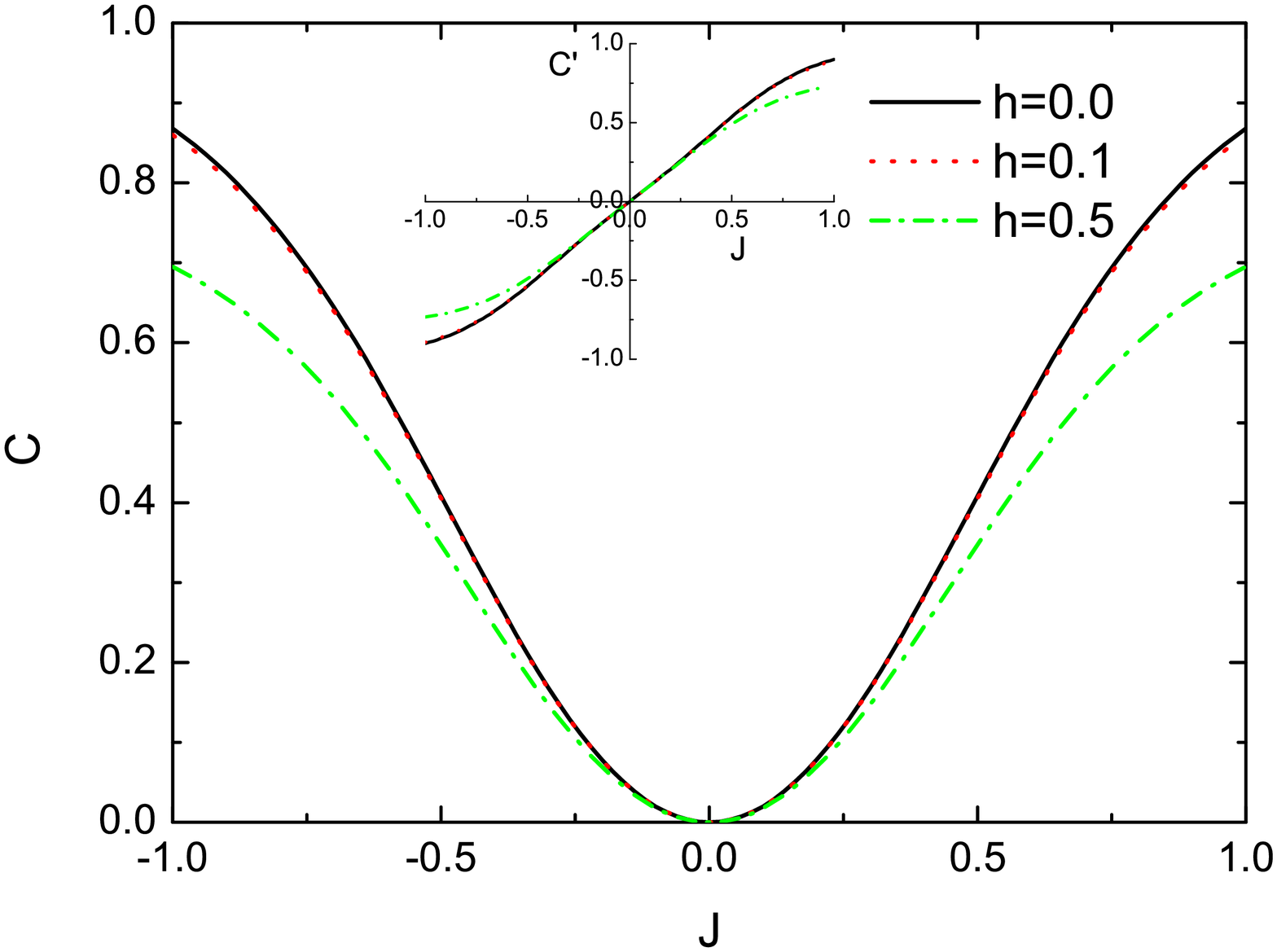}
     \hskip .2cm
  (b)   \includegraphics[bb=82 23 717 519,scale=0.33]{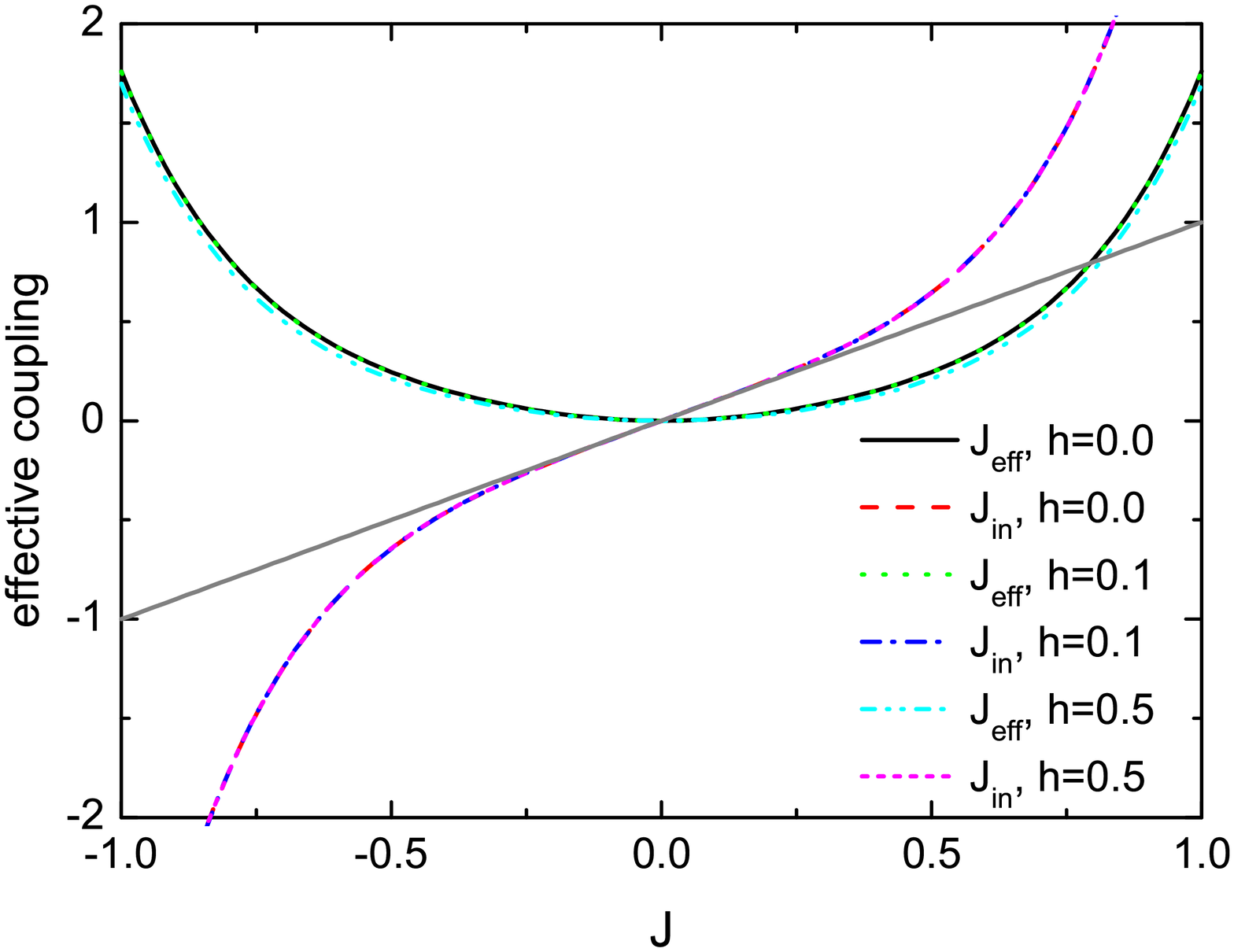}
     \vskip .2cm
     \centering
   (c)  \includegraphics[bb=64 23 714 517,scale=0.4]{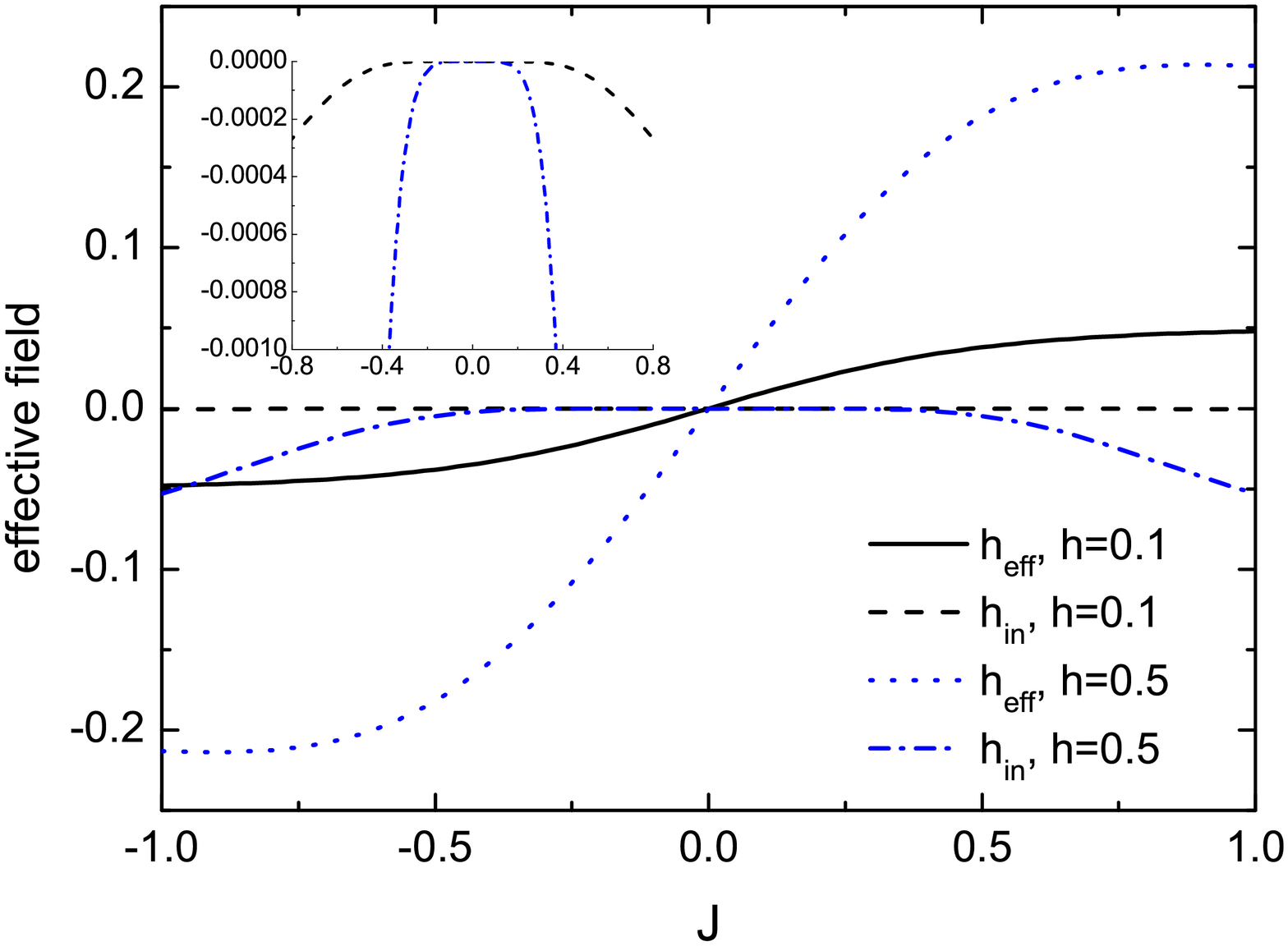}\vskip .2cm
  \caption{(Color online)
     Inference results corresponding to Fig.~\ref{Small} (b). (a)
     connected correlation versus coupling strength between observed
     neurons and the hidden one. The inset shows the connected correlation $C'$ inside the
     observed network. (b) effective coupling between observed
     neurons. The gray line indicates equality. (c) effective field
     for observed neurons. Note that the true value is zero. The
     inset shows an enlarged view.
   }\label{ana02}
 \end{figure}

 The connected correlation $C$ for neurons $1$ and $2$ is shown in
 Fig.~\ref{ana02} (a). Increasing the firing bias of the hidden
 neuron has the effect of lowering the connected correlation, which
 is already observed in a three-neuron system. This also occurs for $C'$ (the inset). For effective coupling,
 the behavior becomes rich. First, the firing bias $h$ affects $J_{{\rm
 eff}}$ in a similar manner to that observed in Fig.~\ref{ana01} (b). However, it
 does not affect $J_{{\rm in}}$, and $J_{{\rm in}}$ is close to its
 true value only when the strength of $J$ is small ($|J|<0.22$). As
 observed in Fig.~\ref{ana02} (c), $h_{{\rm eff}}$ and $h_{{\rm
 in}}$ also show different behavior. The magnitude of $h_{{\rm
 in}}$ is much smaller than that of $h_{{\rm eff}}$, whereas, both
 of them grow with the firing bias $h$. There exists a range of $J$
 where the inferred $h_{{\rm
 in}}$ takes a value close to zero. This range becomes narrow as $h$
 increases. $h_{{\rm eff}}$ as a function of $J$ shows a behavior
 similar to that in the three-neuron system (see Fig.~\ref{ana01} (c)).

\subsection{Two-neuron interaction mediated by a chain of hidden neurons}
\label{subsec:ana03}

 The interaction between two observed neurons can also be mediated
 by a chain of unobserved neurons which interact with each other by a
 coupling strength $J$ (see Fig.~\ref{Small} (c)). Here we assume homogeneous interactions in
 the hidden part and no firing bias for hidden neurons, by focusing
 on how the effective coupling varies with the chain length $L$ and
 the coupling strength $J$. This chain will cause a correlation
 between observed neurons~\cite{SM-1992},
 \begin{equation}\label{case-c}
    C=(\tanh J)^{L}.
 \end{equation}
And correspondingly, an effective coupling is given by
\begin{equation}\label{J-c}
    J_{{\rm eff}}=\frac{(\tanh J)^{L}}{1-(\tanh J)^{2L}}.
 \end{equation}

\begin{figure}[h!]
 (a) \includegraphics[bb=72 18 714 513,scale=0.33]{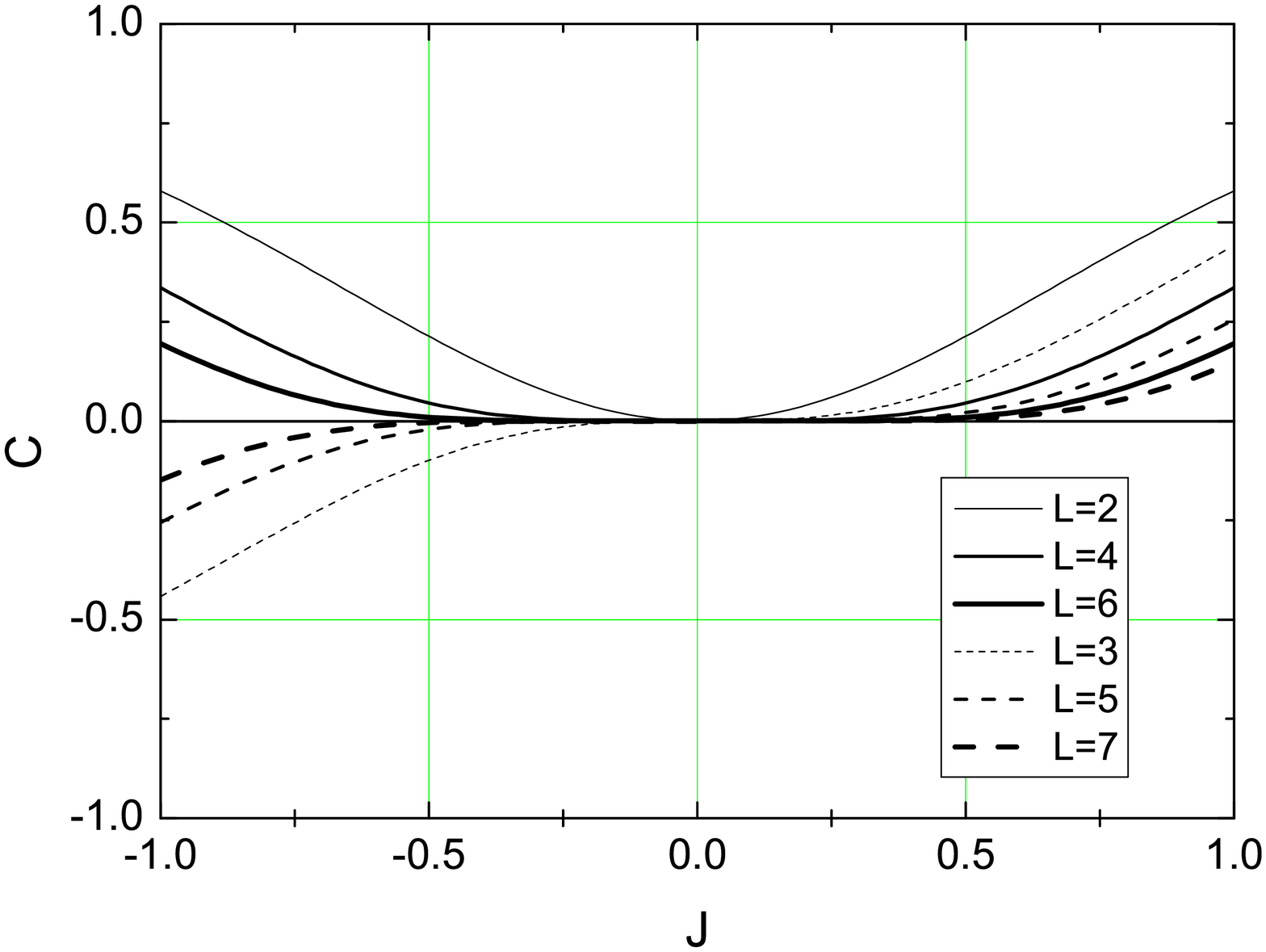}
     \hskip .5cm
  (b)   \includegraphics[bb=73 23 717 519,scale=0.33]{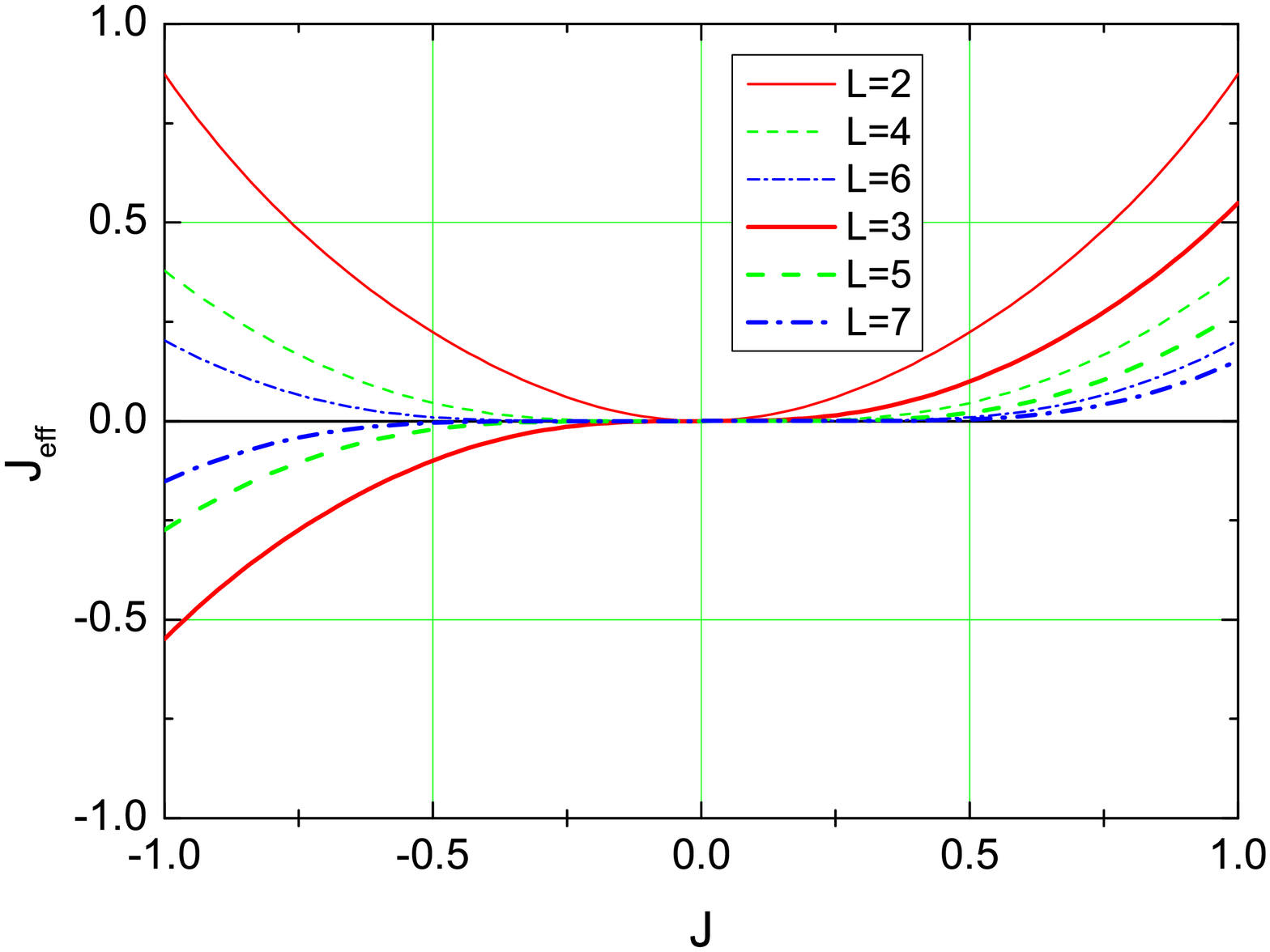}
     \vskip .2cm
  \caption{(Color online)
     Inference results corresponding to Fig.~\ref{Small} (c). (a)
     connected correlation versus coupling strength in the hidden part. (b) effective coupling between observed
     neurons. $L$ is the chain length.
   }\label{ana03}
 \end{figure}

The effect of $J$ and $L$ is shown in Fig.~\ref{ana03}. We find that
the correlation becomes weak as $L$ increases, and the magnitude of the
effective coupling also decreases with increasing $L$. $J_{{\rm
eff}}$ is close to the true zero value only when $J$ falls within
certain interval, otherwise, $|J_{{\rm eff}}|$ grows with $J$. This
is consistent with the small correlation assumption made in the
naive mean field approximation~\cite{Kappen-1998,Tanaka-1998}.

\begin{figure}[h!]
 (a) \includegraphics[bb=66 537 304 692,scale=0.7]{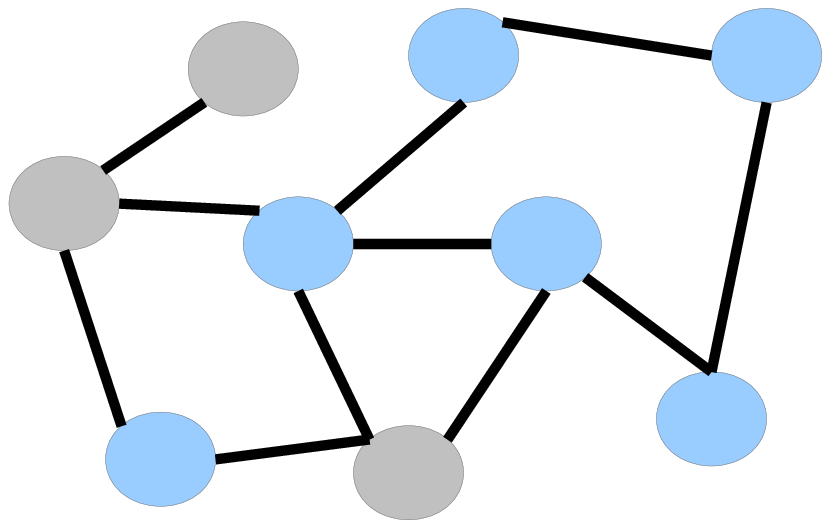}
     \hskip .5cm
  (b)   \includegraphics[bb=53 37 735 536,scale=0.35]{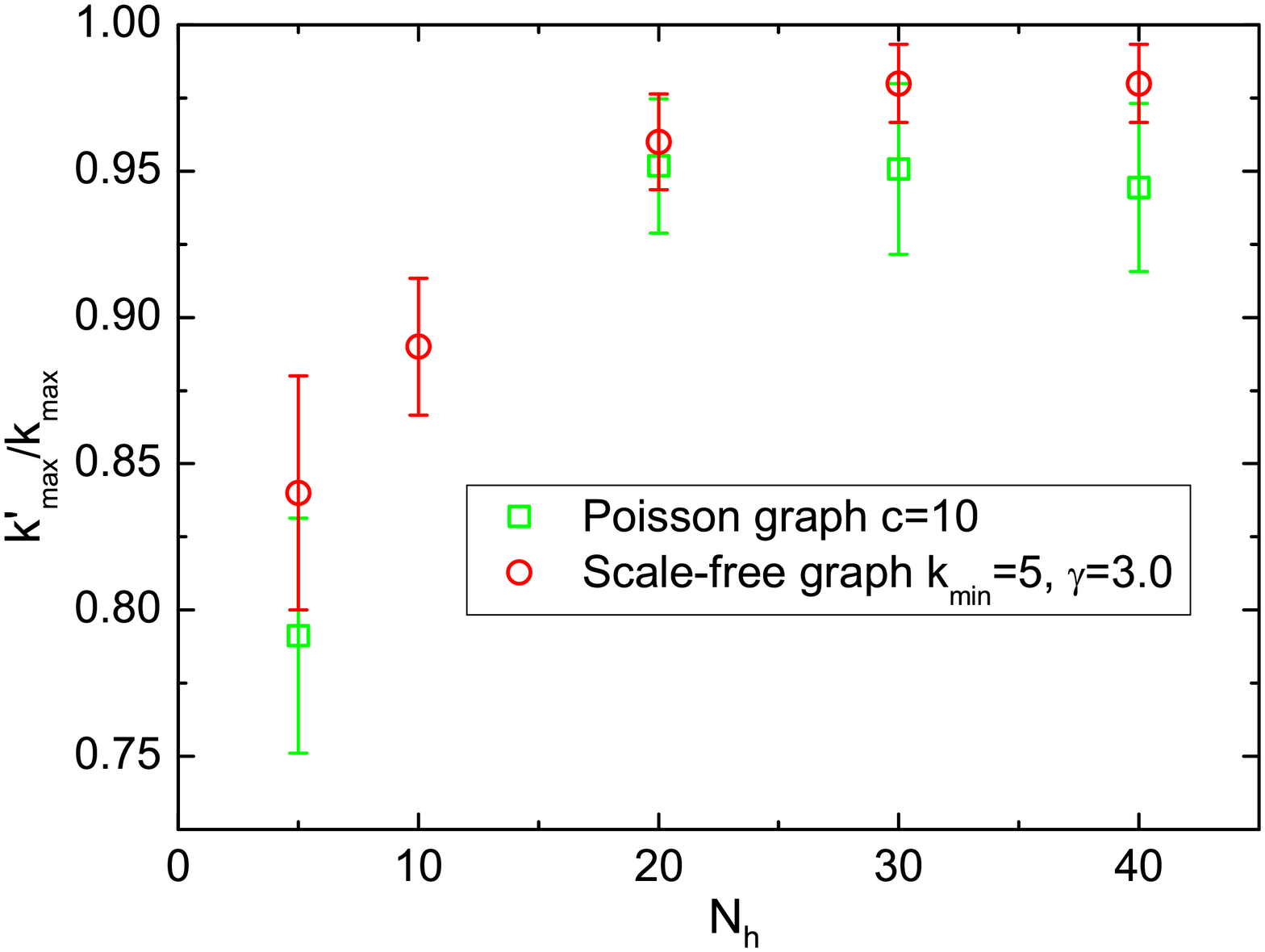}
     \vskip .2cm
  \caption{(Color online)
      (a) Network structure for large size system study. We show an example of $(N,N_h)=(9,3)$. The gray
      circles indicate the hidden nodes while the others are visible
      nodes. In our simulations, we used $N=100$ with increasing value of $N_h$. Nodes are sparsely connected with either Poisson or power-law degree
      distribution. The set of hidden nodes is randomly selected for each network instance.
       (b) The ratio between the maximal degree of hidden nodes and
       that of all nodes. The random Poisson graph has mean degree $c$ and
       the random scale-free graph is characterized by the degree
       distribution $P(k)\propto k^{-\gamma} (k\geq k_{{\rm min}})$.
       The result is averaged over ten random instances.
   }\label{graph}
 \end{figure}
\section{Large size system: numerical simulations} \label{sec:num}
For large size system, it is very difficult to perform an analytic
calculation. However, the main effects of hidden nodes on the
inference quality can be probed by numerical simulations. Here, we
consider a system of size $N=100$, and the mean connectivity of each
node $c=10$ unless otherwise specifically stated. Fig.~\ref{graph}
shows the network structure we consider. We consider the model
defined on random Poisson graphs~\cite{Mezard-1987epl} first, and random scale-free graphs~\cite{Kim-2005}
at the end of this section. As the number of hidden nodes grows, the
nodes of high connectivity (hubs) appear with higher probability in
the hidden part (see Fig.~\ref{graph} (b)). These hubs may play an
important role in affecting the inference quality of the observed
part.

The state space of the model is sampled by a standard Monte-Carlo
procedure, which consists of an asynchronous update of all neurons
at an elementary time step ($N$ proposed neuron's state flips), i.e., the state of neuron $i$, say
$\sigma_i$ is updated by
\begin{equation}\label{MC}
    Prob(\sigma_i\rightarrow-\sigma_i)=\exp\left(-2\sigma_iH_i\right),
\end{equation}
where $H_i=h_i+\sum_{j\neq i}J_{ij}\sigma_j$, and the update goes over all $i$. The experimental data
is collected as $P=10^{5}$ independent configurations (the inference
will become more accurate with larger $P$~\cite{SM-09}); each of
them was sampled with an interval equal to $40$ elementary time
steps after sufficient thermal
equilibration. These data are used to compute correlations and
magnetizations. The inference quality is evaluated by the (relative)
root-mean-square errors:
\begin{eqnarray}\label{rms}
\delta_{J}&=&\left[\frac{\sum_{i<j}(J_{ij}^{VV}-\tilde{J}_{ij}^{VV})^{2}}{\sum_{i<j}(\tilde{J}_{ij}^{VV})^{2}}\right]^{1/2},\\
\Delta
h&=&\left[\frac{\sum_{i}(h_i^{V}-\tilde{h}_i^{V})^{2}}{N-N_{h}}\right]^{1/2}.
\end{eqnarray}
$\tilde{J}_{ij}^{VV}$($\tilde{h}_i^{V}$) represents the true value.
The reported results are the average over ten random realizations of
the network, with the error bars showing the standard deviation.

\subsection{Inference performance on random Poisson graph}
\label{subsec:Pois}
\subsubsection{Inference performance with increasing coupling
variance} \label{subsec:sim01}
\begin{figure}
\centering
    \includegraphics[bb=61 36 719 516,scale=0.4]{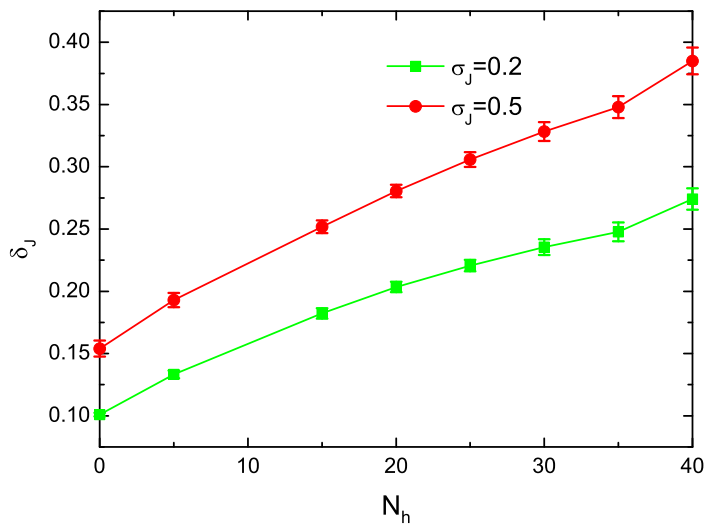}
  \caption{
  (Color online) Inference performance on random Poisson graphs ($c=10$) without firing biases for
  hidden neurons.
  }\label{noh}
\end{figure}

Fig.~\ref{noh} reports results on
networks with $h_{i}^{H}=0$ for any hidden neuron $i$. Increasing
the number of unobserved neurons, one observes that the inference
error $\delta_J$ also increases. One possible reason is, the growing
unobserved part yields larger and larger correlations among observed
neurons, resembling a glassy effect caused by increasing the
coupling variance~\cite{Huang-2013}. These correlations may contain
higher-order ones. The overall effect is to cause some predicted
couplings deviate strongly from their true values. The error
increases with $\sigma_J$ as well, which is consistent with findings
obtained in small systems (see Fig.~\ref{ana01} (b) and
Fig.~\ref{ana02} (b)).

\begin{figure}
\centering
    \includegraphics[bb=61 36 719 516,scale=0.4]{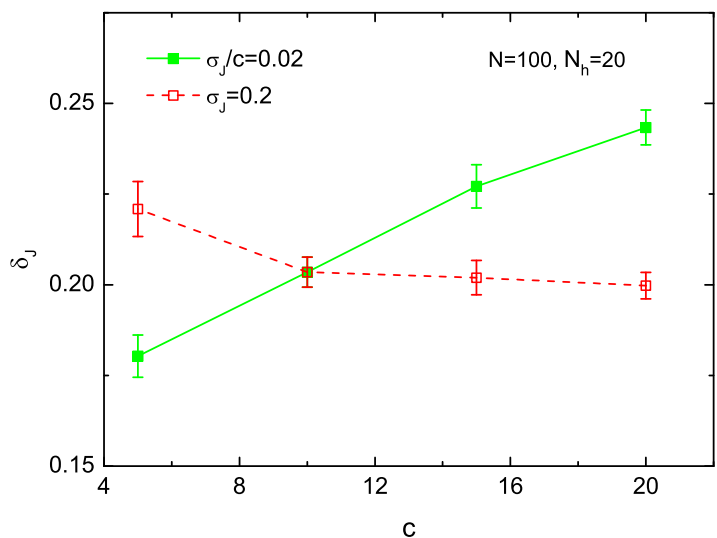}
  \caption{
  (Color online) Inference performance on random Poisson graphs versus $c$ without firing biases for
  hidden neurons.
  }\label{nohC}
\end{figure}

As shown in Fig.~\ref{nohC}, by increasing $c$ but maintaining the
same value of $\sigma_J/c$ , one should observe a larger effect of
the hidden nodes, since this will increase the probability for a
hidden node to have an interaction with the observed nodes. However,
if only $\sigma_J$ is fixed, the error will decrease with $c$
because the overall strength of coupling is weakened.

\subsubsection{Inference performance with increasing field variance
of hidden nodes} \label{subsec:sim02}
\begin{figure}[h!]
 (a) \includegraphics[bb=72 18 714 513,scale=0.33]{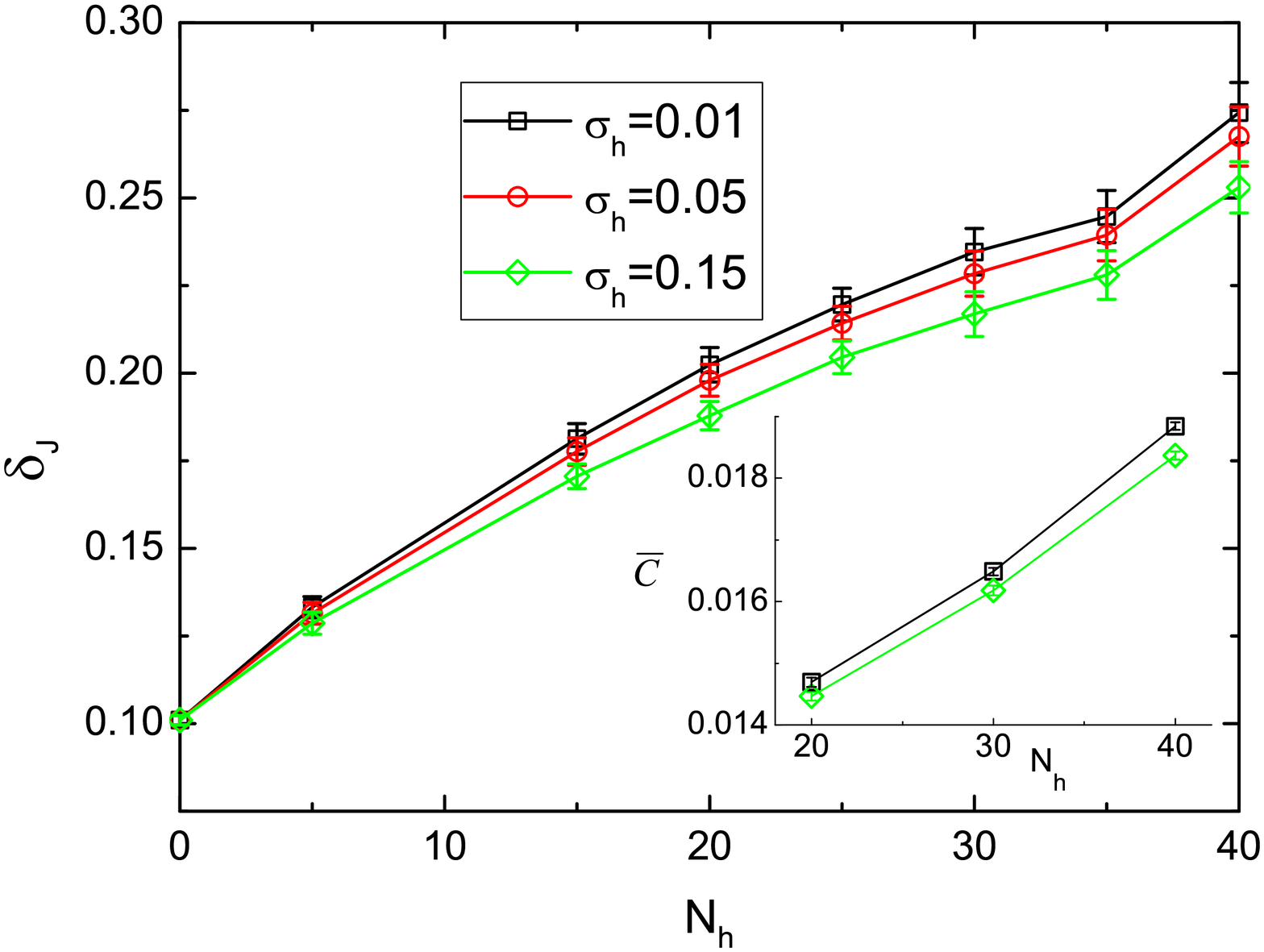}
     \hskip .5cm
  (b)   \includegraphics[bb=73 23 717 519,scale=0.33]{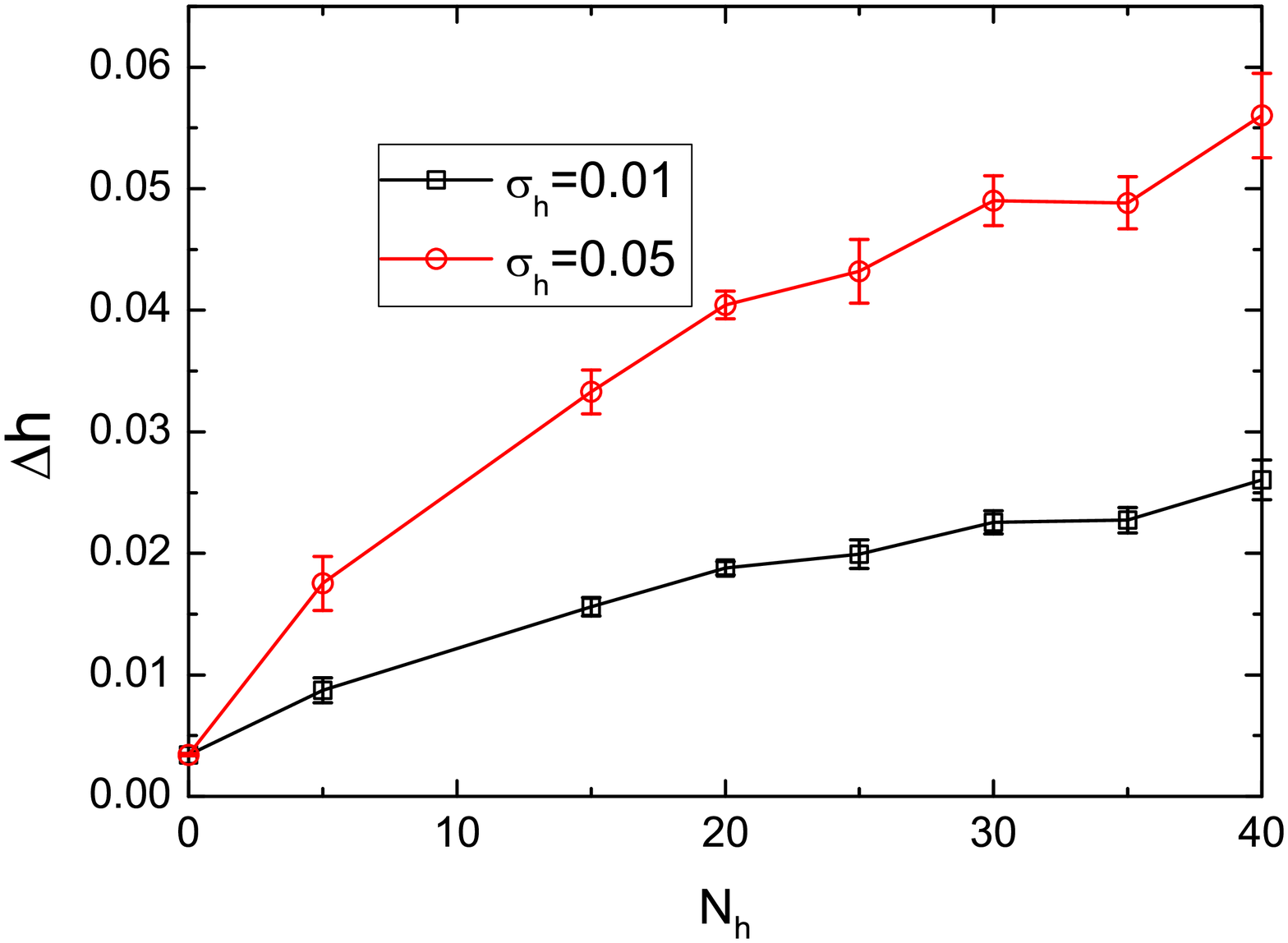}
  \vskip .5cm
     \centering
   (c)  \includegraphics[bb=64 17 737 536,scale=0.4]{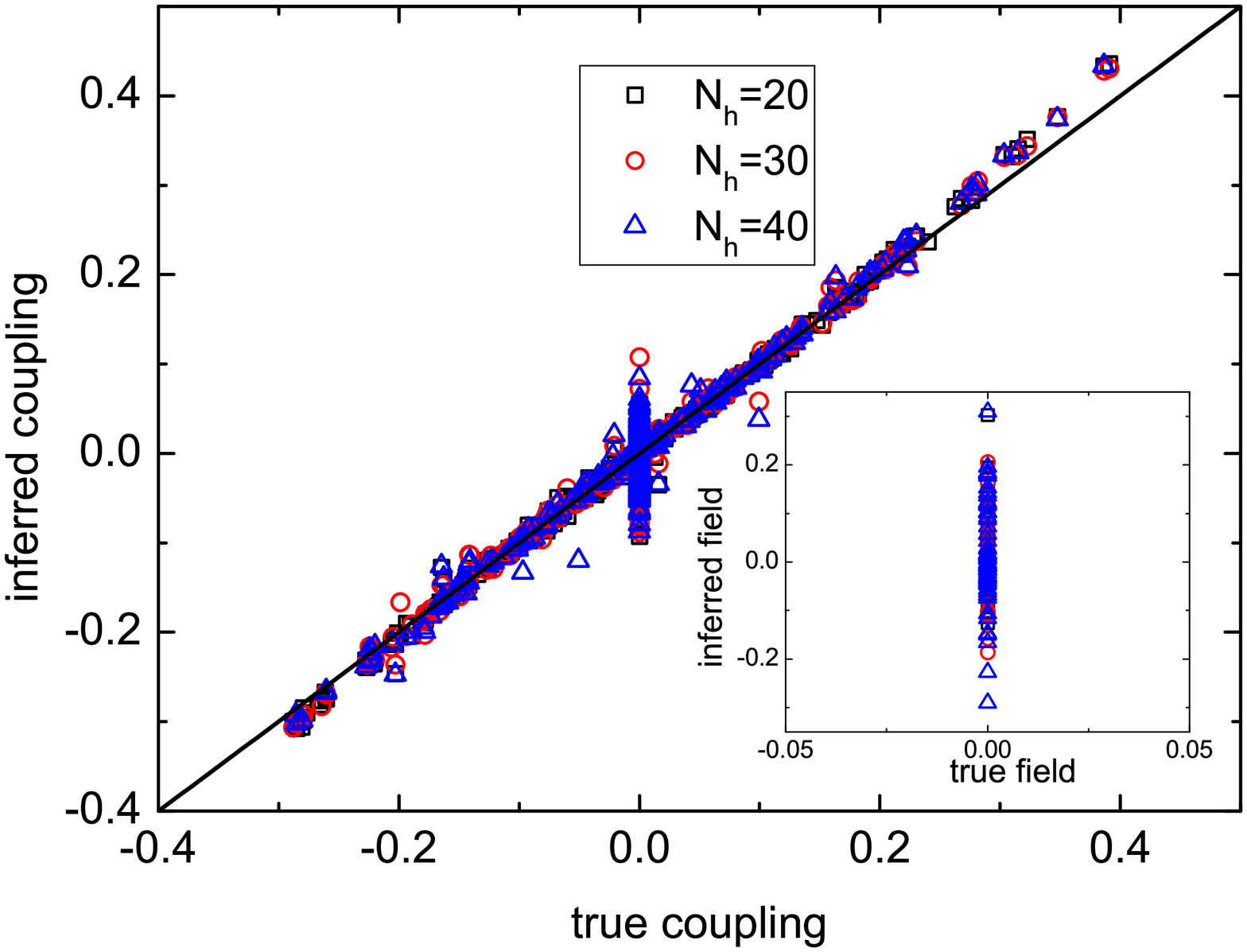}
     \vskip .2cm
  \caption{(Color online)
     Inference performance on random Poisson graphs ($c=10$) with $\sigma_J=0.2$. (a) Inference error for couplings versus the number of hidden
     neurons. The inset gives the global correlation versus the number of hidden neurons (the error bars are smaller than the symbol size). (b) Inference error for fields versus the number of hidden
     neurons. (c) The scatter plot comparing inferred couplings with the true ones for a typical example with
     $\sigma_h=0.15$. The line indicates equality. The inset shows
     the scatter plot for fields.
   }\label{hidh}
 \end{figure}

To explore the effect of the
field strength of hidden nodes, one can keep a small coupling
variance while increasing the value of $\sigma_h$. Interestingly,
the coupling error decreases as $\sigma_h$ increases, which becomes
more apparent when $N_h$ gets larger, as shown in Fig.~\ref{hidh}
(a). This result is consistent with that observed in
Fig.~\ref{ana01} (b) and Fig.~\ref{ana02} (b). This can be explained
by the fact that, increasing the number of unobserved neurons will
make more observed neurons interact directly with the hidden ones,
while the effective couplings between these observed neurons are
expected to give a large contribution to the inference error. In
particular, increasing field strength of hidden nodes results in
smaller correlations, and thus the coupling prediction can be
improved. This point can be easily understood from the analytic
study of small systems, as shown in Fig.~\ref{ana01} (a) and
Fig.~\ref{ana02} (a). Compatible with this effect, we also show the
global mean correlation defined as $\bar{C}=\left[\frac{1}{|\mathcal
{P}_{\rm obs}|}\sum_{(i,j)\in\mathcal {P}_{\rm
obs}}C_{ij}^{2}\right]^{1/2}$ where $\mathcal {P}_{\rm obs}$ denotes
the pair set of observed nodes, in the inset of Fig.~\ref{hidh} (a).
We see clearly the global correlation decreases as the field
variance increases.

In contrast to the coupling error, the field error still grows with
the field variance, which becomes more evident at larger $N_h$, as shown in Fig.~\ref{hidh} (b). This
may be related to the fact that, the effective fields of observed
neurons interacting directly with the hidden neurons yield a larger
contribution to the inference error, compared to those inside the
observed part (not on the boundary between observed and unobserved
part). The effective fields of the boundary neurons seem to be very
sensitive to changes of firing biases of hidden neurons, as observed
in Fig.~\ref{ana01} (c) and Fig.~\ref{ana02} (c).

As shown in
Fig.~\ref{hidh} (c), in the presence of hidden nodes, the inferred coupling values over-estimate
the true large positive couplings, while the large (in absolute value) negative couplings are slightly
under-estimated, which was also observed in similar works~\cite{Roudi-2009,Lezon-2006}. As $N_h$ varies, the inference error is
mainly caused by non-existent connections and those connections with weak couplings.

\subsubsection{Inference performance with re-scaled couplings}
\label{subsec:sim03}
\begin{figure}[h!]
 (a) \includegraphics[bb=72 18 714 513,scale=0.33]{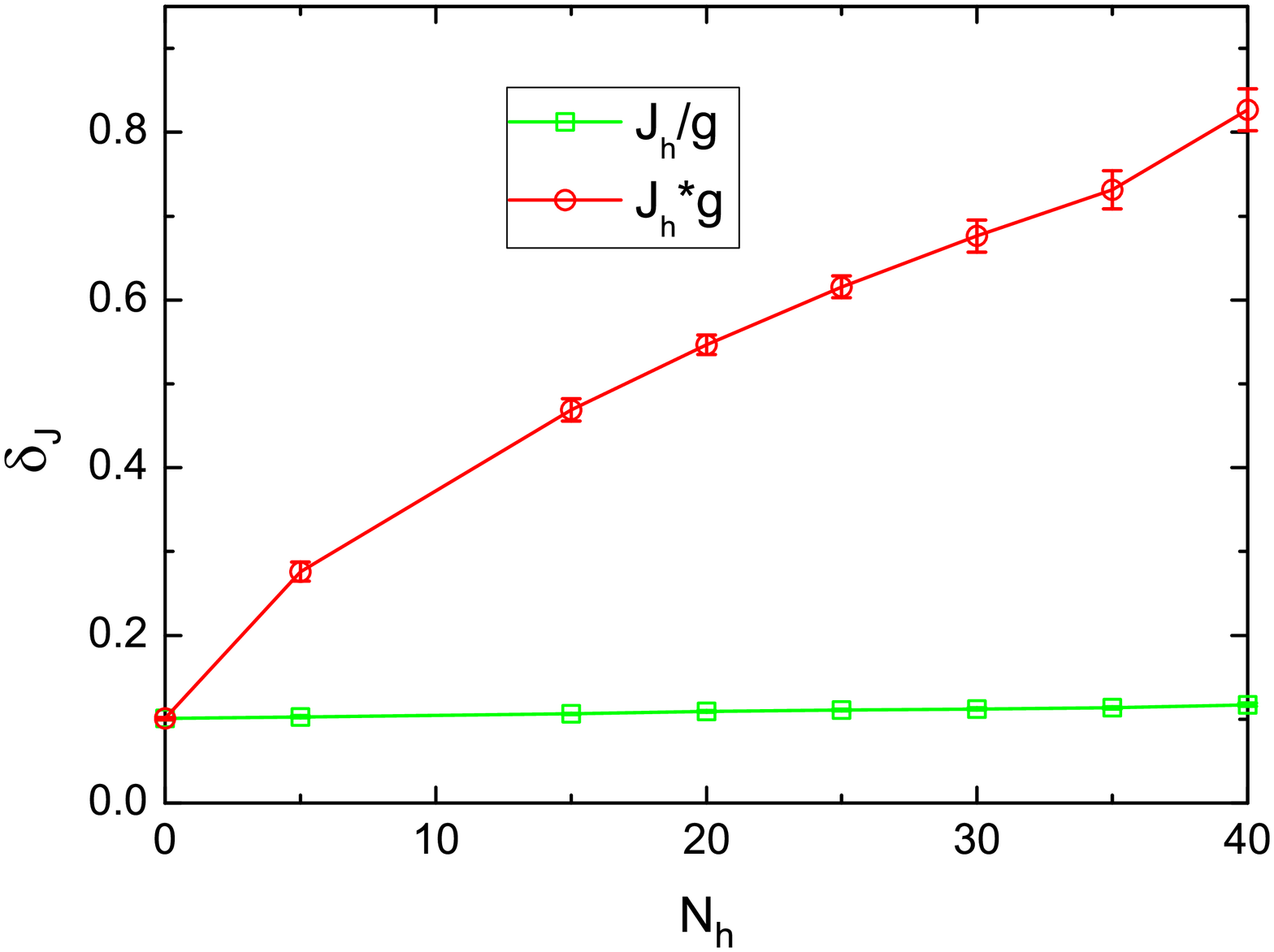}
     \hskip .5cm
  (b)   \includegraphics[bb=73 23 717 519,scale=0.33]{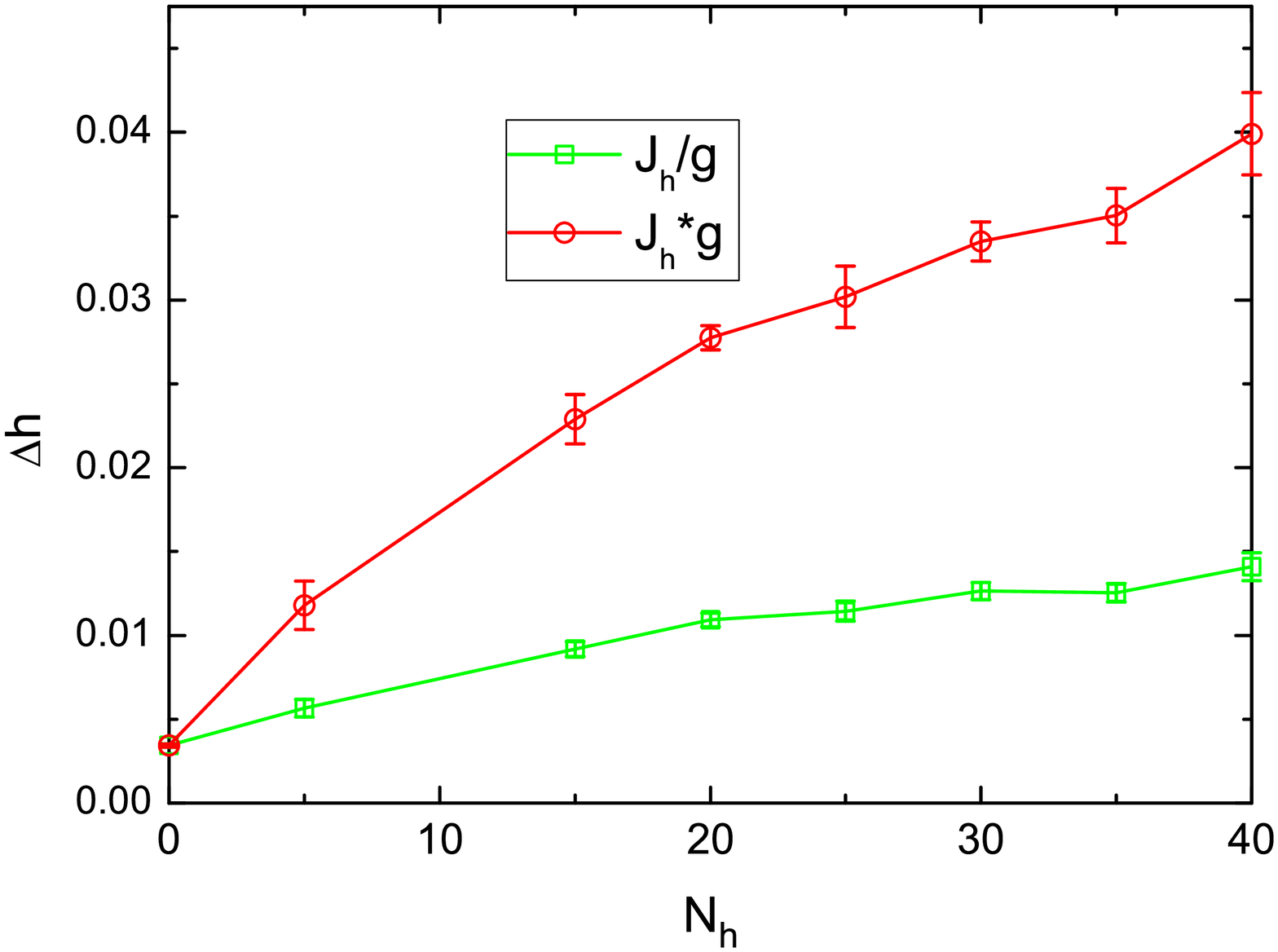}
     \vskip .2cm
  \caption{(Color online)
     Inference performance on random Poisson graphs ($c=10$) with $\sigma_J=0.2$ and $\sigma_h=0.01$. The interaction involved in at least one of
     the hidden neurons is enhanced by a factor $g=2$ or attenuated by a factor $1/g$.  (a) inference error for couplings versus the number of hidden
     neurons. (b) inference error for fields versus the number of hidden
     neurons.
   }\label{hidhJ}
 \end{figure}

Keeping a low value of $\sigma_h$, one can also
explore the effect of coupling strength among hidden neurons or
between hidden and visible neurons, by re-scaling the couplings,
i.e., all $J^{VH}$ and $J^{HH}$ (hidden couplings) are enhanced by a
factor $J\rightarrow gJ$ or attenuated as $J\rightarrow J/g$, where
we choose $g=2$. Results are reported in Fig.~\ref{hidhJ}. This case
corresponds to adding a large perturbation to the couplings related
to the hidden neurons, with the consequence that both the coupling
and the field error increase when hidden couplings are enhanced.
This finding is also consistent with the results reported in small
systems (see Fig.~\ref{ana01} and Fig.~\ref{ana02}).

\subsection{Inference performance on random scale-free graph}
\label{subsec:SF}
\begin{figure}[h!]
 (a) \includegraphics[bb=65 25 735 534,scale=0.33]{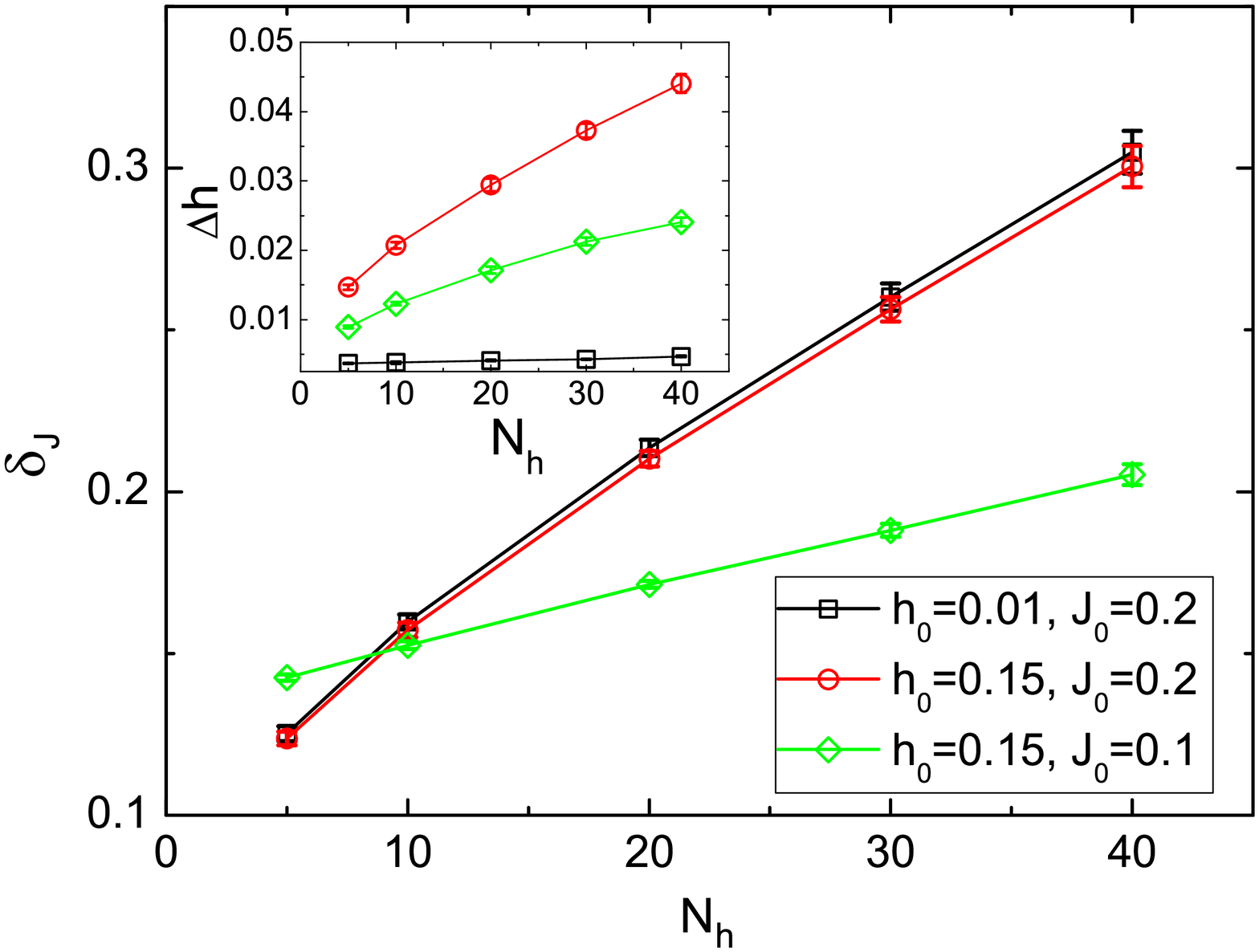}
     \hskip .5cm
  (b)   \includegraphics[bb=59 19 735 539,scale=0.33]{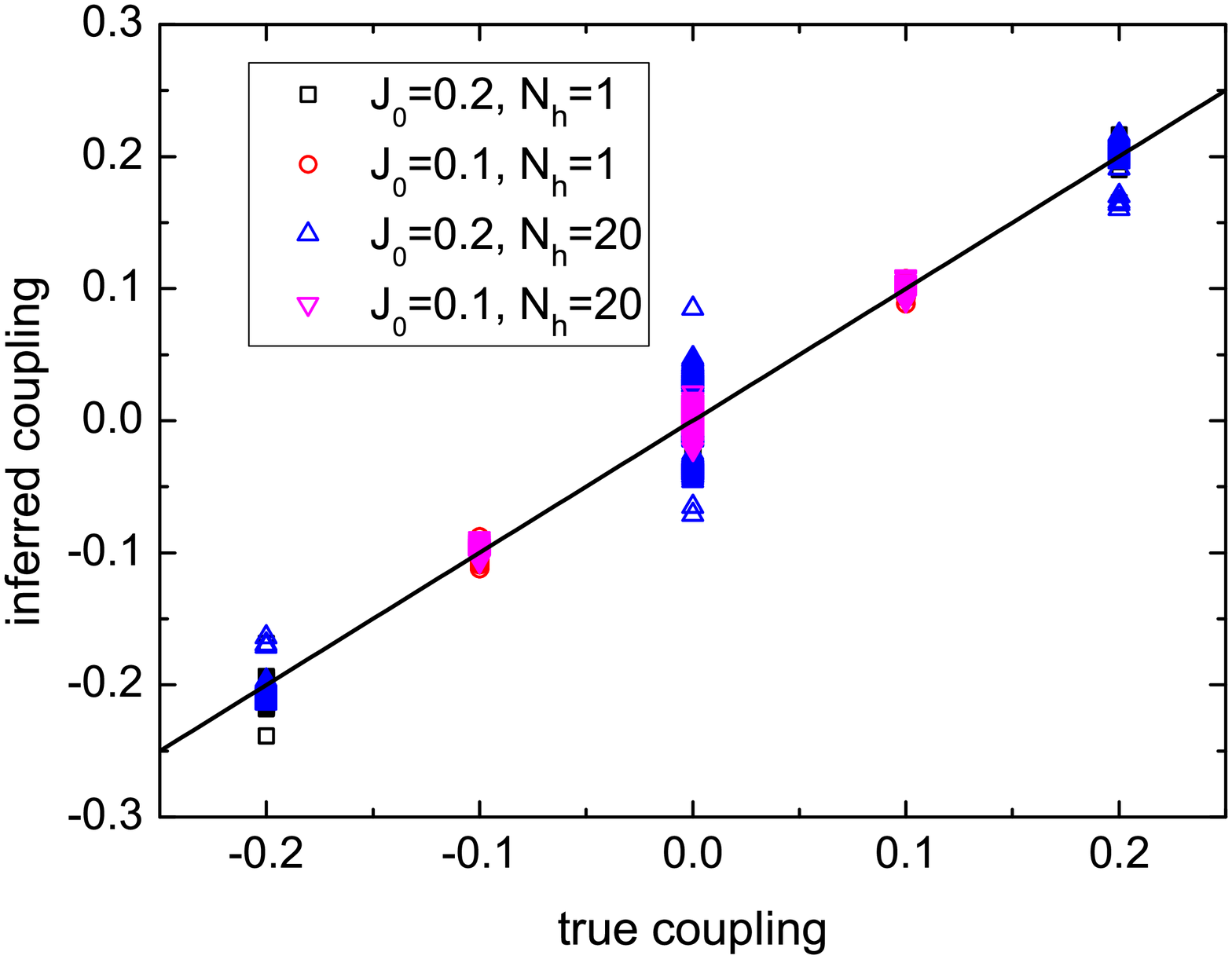}
  \vskip .5cm
     \centering
   (c)  \includegraphics[bb=66 37 736 541,scale=0.4]{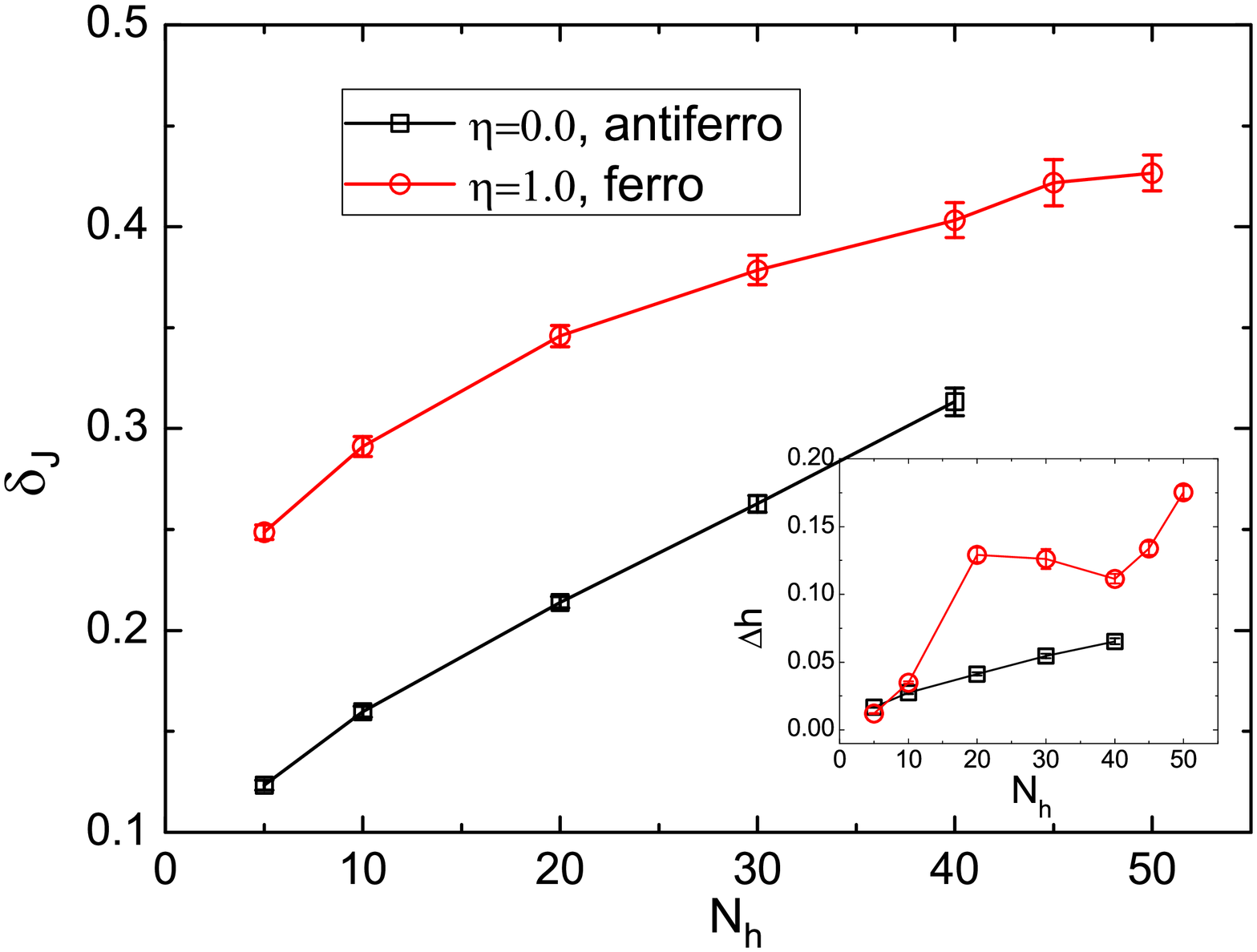}
     \vskip .2cm
  \caption{(Color online)
     Inference performance on random scale-free graph ($k_{{\rm min}}=5,\gamma=3$). (a) Inference error versus the number of hidden
     neurons with $\eta=0.5$. (b) The scatter plot for a typical example with $\eta=0.5$ and
     $h_0=0.15$. The line indicates equality.
     (c) Inference error for ferromagnetic and anti-ferromagnetic
     model with $J_0=0.2$ and $h_0=0.15$.
   }\label{SF}
 \end{figure}

In this section, we study the effects of hidden
nodes on random scale-free graph. Here the coupling strength follows
the binary distribution
$p(J)=\eta\delta(J-J_0)+(1-\eta)\delta(J+J_0)$ and the field is kept
constant ($h=h_0$). In this case, $\eta=1.0$ corresponds to the
ferromagnetic model while $\eta=0.0$ corresponds to the
anti-ferromagnetic model. The behavior Fig.~\ref{SF} (a) shows is similar
to that observed in Fig.~\ref{noh} and Fig.~\ref{hidh}. Note
that when $N_h$ becomes small, the relative error $\delta_J$ shows
larger value at smaller $J_0$. This is because the overall strength
of the denominator in the definition of $\delta_J$ dominates the
error when $N_h$ is small, and does not mean that the inference
quality at the large $J_0$ is better than that at the small $J_0$
(see Fig.~\ref{SF} (b) for the scatter plots of a typical example).
Fig.~\ref{SF} (c) shows the inference performance of ferromagnetic
and anti-ferromagnetic models, implying that the inference quality
of either coupling or field deteriorates as the number of hidden
nodes increases.

\section{Conclusion}
\label{sec_con} In this work, we study effects of unobserved part
of a network on the inference quality of network structure of the
observed part. We first study analytically the small size network
with a few neurons, and find that, the effective couplings between
the boundary neurons decreases as the firing bias of the hidden
common input neuron increases, whereas, this firing bias does not
affect the effective couplings inside the observed network. These
effective couplings grow as the magnitude of the input coupling
strengthes increases. Increasing firing bias of the hidden input
will also increase the effective field of the boundary neurons which
show different behavior from those neurons inside the observed part.

All these interesting properties are also observed in numerical
simulations of large networks. The inference quality for both
couplings and fields deteriorates as the size of the unobserved part
grows, which can be explained by the fact that an increasing number
of hidden neurons causes higher-order correlations in the observed
data and furthermore a larger deviation of the inferred coupling
from its true value, mimicking a glassy phase arising in observed
networks. Interestingly, increasing field variance of hidden neurons
improves the inference quality of the effective couplings, but
worsens the quality for the effective fields. In addition,
attenuated coupling strengthes involved in at least one hidden
neuron lead to high quality of coupling inference. Our work
demonstrates the hidden part in a full network does have a
significant influence on the inference quality of the observed
network structure (for both Poisson graph and scale-free graph),
showing many interesting properties, as revealed in both small
networks and large networks.

As new advanced experimental recording techniques are proposed, the
number of measured neurons becomes larger and larger, providing
greater challenges for large-scale neural data analysis. The effects
of hidden neurons on the inference quality, based on a toy model
study in this paper, provide insights towards understanding the
interaction between observed part and unobserved part in terms of
network structure prediction. Another interesting extension of the
current work is to consider the Potts model which is widely used in
protein structure prediction~\cite{Monasson-2014} (and references
therein).


\section*{Acknowledgments}

I thank two anonymous referees for their constructive comments. This
work was partially supported by the JSPS Fellowship for Foreign
Researchers Grant No. $24\cdot02049$ and partially supported by the program for Brain Mapping by Integrated Neurotechnologies for Dis
ease Studies (Brain/MINDS) from Japan Agency for Medical Research and development, AMED.

\section*{References}
\bibliographystyle{unsrt}

\begin{thebibliography}{10}

\bibitem{Marre-2012}
O.~Marre, D.~Amodei, N.~Deshmukh, K.~Sadeghi, F.~Soo, T.~E. Holy, and M.~J.
  Berry.
\newblock {\em J. Neurosci.}, 32:14859--14873, 2012.

\bibitem{Buzaki-2014}
A.~Ber{\'e}nyi, Z.~Somogyvari, A.~J. Nagy, L.~Roux, J.~D. Long, S.~Fujisawa,
  E.~Stark, A.~Leonardo, T.~D. Harris, and G.~Buzsaki.
\newblock {\em J. Neurophysiol}, 111:1132--1149, 2014.

\bibitem{Yu-2014}
J.~P. Cunningham and B.~M. Yu.
\newblock {\em Nat. Neurosci.}, 17:1500--1509, 2014.

\bibitem{Rieke-2008}
P.~K. Trong and F.~Rieke.
\newblock {\em Nat. Neurosci.}, 11:1343--1351, 2008.

\bibitem{Hertz-11}
J.~Hertz, Y.~Roudi, and J.~Tyrcha.
\newblock e-print arXiv:1106.1752, 2011.

\bibitem{Tkacik-2014}
G.~Tkacik, O.~Marre, D.~Amodei, E.~Schneidman, W.~Bialek, and M.~J.~Berry II.
\newblock {\em PLoS Comput Biol}, 10:e1003408, 2014.

\bibitem{Paninski-13}
S.~{Keshri}, E.~{Pnevmatikakis}, A.~{Pakman}, B.~{Shababo}, and L.~{Paninski}.
\newblock e-print arXiv:1309.3724, 2013.

\bibitem{Friedman-1998}
Nir Friedman, Kevin Murphy, and Stuart Russell.
\newblock In {\em Proceedings of the Fourteenth Conference on Uncertainty in
  Artificial Intelligence}, UAI'98, pages 139--147, San Francisco, CA, USA,
  1998. Morgan Kaufmann Publishers Inc.

\bibitem{Roudi-2013}
B.~Dunn and Y.~Roudi.
\newblock {\em Phys. Rev. E}, 87:022127, 2013.

\bibitem{Hertz-2014}
J.~Tyrcha and J.~Hertz.
\newblock {\em Mathematical Biosciences and Engineering}, 11:149--165, 2014.

\bibitem{Opper-2014}
L.~Bachschmid-Romano and M.~Opper.
\newblock {\em J. Stat. Mech.: Theory Exp}, 2014(6):P06013, 2014.

\bibitem{Nature-06}
E.~Schneidman, M.~J. Berry, R.~Segev, and W.~Bialek.
\newblock {\em Nature}, 440:1007, 2006.

\bibitem{Huang-2012pre}
H.~Huang and H.~Zhou.
\newblock {\em Phys. Rev. E}, 85:026118, 2012.

\bibitem{Cocco-13}
J.~Barton and S.~Cocco.
\newblock {\em J. Stat. Mech.: Theory Exp}, page P03002, 2013.

\bibitem{Kappen-1998}
H.~J. Kappen and F.~B. Rodriguez.
\newblock {\em Neural Comput}, 10:1137, 1998.

\bibitem{Tanaka-1998}
T.~Tanaka.
\newblock {\em Phys. Rev. E}, 58:2302, 1998.

\bibitem{Huang-2013}
H.~Huang and Y.~Kabashima.
\newblock {\em Phys. Rev. E}, 87:062129, 2013.

\bibitem{Jaynes-1957}
E.~T. Jaynes.
\newblock {\em Phys. Rev.}, 106:620--630, 1957.

\bibitem{Mezard-1987}
M.~M\'ezard, G.~Parisi, and M.~A. Virasoro.
\newblock {\em Spin Glass Theory and Beyond}.
\newblock World Scientific, Singapore, 1987.

\bibitem{SM-1992}
J.~M. Yeomans.
\newblock {\em Statistical Mechanics of Phase Transitions}.
\newblock Oxford University Press, Oxford, 1992.

\bibitem{Ack-1985}
D.~H. Ackley, G.~E. Hinton, and T.~J. Sejnowski.
\newblock {\em Cognitive Science}, 9:147, 1985.

\bibitem{Saul-1996}
Lawrence~K. Saul, Tommi Jaakkola, and Michael~I. Jordan.
\newblock {\em Journal of Artificial Intelligence Research}, 4:61--76, 1996.

\bibitem{Hinton-2006}
G~Hinton, S~Osindero, and Y~Teh.
\newblock {\em Neural Computation}, 18:1527--1554, 2006.

\bibitem{Mezard-1987epl}
M.~M\'ezard and G.~Parisi.
\newblock {\em EPL (Europhysics Letters)}, 3:1067, 1987.

\bibitem{Kim-2005}
D.-H. Kim, G.~J. Rodgers, B.~Kahng, and D.~Kim.
\newblock {\em Phys. Rev. E}, 71:056115, 2005.

\bibitem{SM-09}
V.~Sessak and R.~Monasson.
\newblock {\em J. Phys. A}, 42:055001, 2009.

\bibitem{Roudi-2009}
Yasser Roudi, Joanna Tyrcha, and John Hertz.
\newblock {\em Phys. Rev. E}, 79:051915, 2009.

\bibitem{Lezon-2006}
Timothy~R. Lezon, Jayanth~R. Banavar, Marek Cieplak, Amos Maritan, and Nina~V.
  Fedoroff.
\newblock {\em Proceedings of the National Academy of Sciences of the United
  States of America}, 103:19033--19038, 2006.

\bibitem{Monasson-2014}
J.~P. Barton, S.~Cocco, E.~De~Leonardis, and R.~Monasson.
\newblock {\em Phys. Rev. E}, 90:012132, 2014.

\end{thebibliography}


\end{document}